\documentclass{article}
\usepackage[utf8]{inputenc}
\usepackage{makecell}
\usepackage{settings}
\usepackage{tabularx}
\usepackage{changepage}
\usepackage{arydshln}
\usepackage[official]{eurosym}
\usepackage{svg}

\newcolumntype{B}{>{\centering\arraybackslash\hsize=.09\hsize}X}
\newcolumntype{K}{>{\centering\arraybackslash\hsize=.12\hsize}X}
\newcolumntype{Y}{>{\centering\arraybackslash}X}

\usepackage{float}
\newfloat{Table}{H}{grf}

\usepackage{microtype}              
\usepackage{ragged2e}
\usepackage{tabularray}
\UseTblrLibrary{booktabs}           
\NewTableCommand\category[1][0pt]{
        \SetRow{abovesep+=#1}
        \SetCell[c=4]{l, font=\small\itshape\bfseries}
\SetTblrStyle{contfoot-text}{font=\footnotesize\itshape}}

\usepackage{xargs}
\newcommandx{\transmat}[2][1={i,j},2={}]{\mathcal{P}_{#1}^{#2}}
\newcommandx{\transmatf}[2][1={i,j},2={}]{\tilde{\mathcal{P}}_{#1}^{#2}}

\newcommandx{\transmatbinary}[2][1={i,j},2={}]{\mathcal{M}_{#1}^{#2}}
\newcommandx{\fitness}[2][1={(n)},2={i}]{F^{#1}_#2}
\newcommandx{\fitnesstemp}[2][1={(n)},2={i}]{\tilde{F}^{#1}_#2}
\newcommandx{\complexity}[2][1={(n)},2={j}]{Q^{#1}_#2}              
\newcommandx{\complexitytemp}[2][1={(n)},2={j}]{\tilde{Q}^{#1}_#2}
\newcommandx{\threshacc}{\Theta^{A}}
\newcommandx{\threshtrans}{\Theta^{T}}

\definecolor{dataColor}{rgb}{0.094, 0.631, 0.466}
\definecolor{estimatedColor}{rgb}{0.043, 0.454, 0.71}
\definecolor{calibratedColor}{rgb}{0.835, 0.384, 0.16}

\title{\LARGE
\textbf{Mitigating Farmland Biodiversity Loss \\ \Large A Bio-Economic Model of Land Consolidation and Pesticide Use}
}

\author{\normalsize Elia Moretti$^{1,2}$ and Michael Benzaquen$^{1,2,3}$}

\date{
\small
\textit{
$^1$Chair of Econophysics and Complex Systems,  Ecole Polytechnique,  91128 Palaiseau, France\\%
$^2$LadHyX, UMR CNRS 7646, Ecole Polytechnique, Institut Polytechnique de Paris, 91128 Palaiseau, France\\
$^3$Capital Fund Management, 23 rue de l'Université, 75007 Paris, France\\%
}\medskip
\today}

\begin{document}
\maketitle

\captionsetup{margin=10pt,font=footnotesize,labelfont=bf,labelsep=endash,justification=centerlast}

\begin{abstract}
Biodiversity loss driven by agricultural intensification is a pressing global issue, with significant implications for ecosystem stability and human well-being. {Existing policy instruments have so far proven insufficient in halting this decline, which raises the need to explore the possible feedback loops that are pivotal to ecosystem degradation.
We design a minimal integrated bio-economic agent-based model to qualitatively explore macro-level biodiversity trends, as influenced by individual farmer behavior within simple decision-making processes.} Our model predicts further biodiversity decline under a business-as-usual scenario, primarily due to intensified land consolidation. We evaluate two policy options: reducing pesticide use and subsidizing small farmers. While pesticide reduction rapidly benefits biodiversity in the beginning, it eventually leads to increased land consolidation and  further biodiversity loss. In contrast, subsidizing small farmers by reallocating a small fraction of existing subsidies, stabilizes farm sizes and enhances biodiversity in the long run. The most effective strategy results from combining both policies, leveraging pesticide reduction alongside targeted subsidies to balance economic pressures and consistently improve biodiversity. 
\end{abstract}

\section{Introduction}


In recent decades, Europe has witnessed a dramatic reduction in biodiversity~\cite{ipbes_report_bio_decline}, with significant declines in both the overall biomass~\cite{Species_biomass_decline} and the number of different species~\cite{butterflies_decline}. Likewise, the number of species at risk of extinction has risen~\cite{species_threatened}. 
The decline is particularly severe in farmlands, where bird populations have shrunk~\cite{bird_and_agri}, the number of pollinators has dwindled~\cite{pollinators_decline}, and invasive species are increasingly infiltrating agro-ecosystems~\cite{Invasive_species}. Modern agriculture is widely recognized as the primary driver of this trend~\cite{agri_drive_bio_loss,agri_drive_bio_loss_2}.

Agriculture covers nearly a third of Europe's land area (and about half of France), making it the predominant land use. Rising food demand has driven agricultural intensification~\cite{agri_demand_intensification}, which relies heavily on chemical inputs like fertilizers and pesticides. These chemicals disrupt ecological balances, harming biodiversity and leading to population declines~\cite{fertilizer_pollution,pesticide_biodiversity,pesticide_biodiversity_2}. Additionally, international competition has prompted significant land consolidation~\cite{land_consolidation,land_consolidation_2}, merging smaller farms into larger, uniform fields to boost efficiency. While this transformation benefits large-scale farming, it adversely affects biodiversity by removing diverse habitats such as hedgerows, ditches, and grass strips with wild vegetation, which are crucial for supporting various species~\cite{consolidation_biodiversity}.

Biodiversity plays a crucial role in the functioning of our ecosystem~\cite{bio_ecosystem_wellbeing,bio_ecosystem_wellbeing_2}, underpinning essential services for societal well-being such as water provision and air purification. {Consequently, over the past three decades, numerous public policies have been designed to mitigate agriculture's adverse impacts on biodiversity~\cite{conservation_poliocy_EU}. Often, these policies have implemented agri-environment schemes that offer financial incentives to farmers who adopt environmentally sustainable practices~\cite{PES_EU}. Despite their longstanding application, clear large-scale evidence of these instruments' effectiveness remains elusive~\cite{PES_failure}, leaving the challenge of identifying effective policy solutions to address biodiversity loss unresolved.}

{In response to this challenge, we have developed an integrated bio-economic agent-based model that allows for the exploration of policy scenarios across interdisciplinary boundaries. By integrating ecological and economic insights, our model offers dynamic projections informed by historical data from the French agricultural sector. Our findings clearly show that continuing with current agricultural practices will further exacerbate biodiversity decline. However, we also delineate a promising approach, illustrating how a thoughtfully balanced policy strategy—integrating pesticide reduction with subsidies aimed at supporting small farmers—can halt the ongoing biodiversity decline.}

\section{{Literature Review}}

{Extensive efforts have been made by ecologists and economists to devise effective policy interventions aimed at curbing ongoing environmental degradation. Across these fields, a variety of models have been developed to assess potential policy scenarios. While many models stem predominantly from either ecological or economic disciplines, a holistic approach that integrates both perspectives is widely recognized as essential for effectively addressing agro-environmental sustainability~\cite{multidisciplinary_needed, multidisciplinary_needed_2}. Concentrating on such integrated models, we can categorize them into two broad approaches: (i) static models and (ii) dynamic models.}

Static models primarily aim to quantify the monetary value of ecological services like pollination, pest control, and soil fertility~\cite{biodiversity_price,biodiversity_price_2}. In the vein of cost-benefit analyses~\cite{cost-benefit_biodiversity}, these models typically assess the trade-offs between agricultural practices and the ecological impact associated with different land-use decisions. However, by their very nature, static models are bound to overlook crucial dynamics of agro-ecological systems, which exist in a continually changing, non-stationary environment~\cite{Cost-benefit_critique}. For example, one should account for resource scarcity~\cite{scarcity_in_eco_price}; vital elements like water and nutrients become less abundant and more valuable over time~\cite{Water_scarcity}. Additionally, one should also consider ecological tipping points~\cite{ecosystem_tippingpoing}, where gradual changes can suddenly trigger dramatic shifts in the  ecosystems. The limitations of static models can lead to oversimplified views and inaccurate predictions, hindering the development of optimal policies~\cite{bio_price_critique}.

{Dynamic models, on the other hand, aim to describe the temporal and spatial evolution of agricultural landscapes, capturing the interactions among their various components. Within this category,} the usual debate then arises between standard economic models, which rely on representative profit-optimizing agents, and agent-based models (ABMs) that are more behaviorally oriented. 

Standard economic models are primarily aimed at evaluating worldwide trade policies~\cite{landsparing_sharing_big_model} to determine the optimal allocation of agriculture for enhancing biodiversity, as well as national-level land-use policies~\cite{Mouysset2011}.  Considering agricultural production at the aggregate level, these models use the concept of ``representative farms" to capture the behavior and characteristics of a group of farms. As an example, Mouysset \textit{et al.}~\cite{Mouysset2011} developed a bio-economic model to study sustainable biodiversity management,  combining the community dynamics of 34 bird species with a representative farmer who makes decisions about land use over time. Such models face criticism for relying on a single representative agent, whereas the real agricultural landscape consists of a diverse mosaic of agents shaped by their interactions and behavioral biases~\cite{farmers_interaction,farmers_bias,agrilove}.

{In this context, agent-based models (ABMs) emerge as more promising candidates~\cite{ABM_ecological-economics}. Indeed, by representing farmers as interacting autonomous agents, each with unique goals, resources, and constraints, ABMs can effectively capture the diversity of agricultural practices and their impact on land use and biodiversity~\cite{farmers_interaction, farmers_interaction_2}. This capability makes ABMs particularly suitable for evaluating the dynamics of agro-ecological systems, where understanding the variability in farmers' behaviors is crucial~\cite{ABM_water_management, abm_human_nature_interactions, sustainable_farming_decision}.}

{The literature on agricultural ABMs has expanded rapidly, encompassing diverse areas such as (i) land use and management~\cite{abm_land_use, agri_ABM_review}, (ii) economic and market dynamics~\cite{abm_supply_chain, review_ABM_farming_decision}, (iii) technological adoption and innovation~\cite{abm_agri_sustainability, agrilove}, (iv) social and behavioral aspects~\cite{abm_water_agri_review, ABM_water_management}, and (v) environmental and ecological interactions~\cite{ABM_ecological-economics,sustainable_farming_decision, pampas_bio_ABM, agri_ABM_review}. Of particular interest to our study is the latter category~\cite{farmers_interaction_2}. Within this ABM literature, a variety of decision-making models are employed~\cite{review_ABM_farming_decision, abm_human_nature_interactions, editor_suggestion_2}, ranging from microeconomic rules~\cite{Agripolis}, where agents maximize profit or utility, to psychosocial and cognitive models~\cite{ABM_water_management}, where decisions rely on cognitive abilities or intentions, and heuristic rules~\cite{agrilove, abm_heuristic_weed}, where agents decision is based on simple adaptive rules.}

{While microeconomic and psychosocial/cognitive models can serve as robust tools for quantitative policy exploration due to their realism in decision-making elements, they also come with challenges of their own. Their detailed granular nature often limits their spatial and temporal resolution to a few years and square kilometers~\cite{Agripolis, pampas_bio_ABM}. Then, calibrating these types of ABMs is consistently challenging, as it requires comprehensive data that are often scarce at these scales~\cite{ABM_criticism}. Moreover, biodiversity decline in agricultural landscapes occurs over long timeframes – typically at least 10 years – and is a pan-European phenomenon that emerges clearly at the macroscopic level, regardless of national differences in specific farming systems and regional economic conditions~\cite{ipbes_report_bio_decline, species_threatened, Species_biomass_decline}.}

{In this regard, the use of simple heuristic rules for describing farmer decision-making appears to be an adequate  approach. While the simplicity of this choice does not enable precise quantitative forecasts for real-world scenarios, we believe it may offer a comprehensive understanding of the feedback loops that are pivotal to ecosystem degradation. This approach aligns with the methodological manifesto proposed by Bouchaud, who advocates for the use of ABMs as explanatory tools to ascertain the plausibility of the existence of ``Black Swans" or ``Dark Corners" – discontinuity lines beyond which runaway instabilities may emerge~\cite{JP_economic_complexity_manifesto, mark0}. Furthermore, this type of model allows for the extension of the simulation horizon both spatially and temporally, leading to a macroscopic description where aggregate properties exhibit minimal fluctuations, thereby making the results more robust against changes in microscopic details. }

{We draw from this extensive body of literature to propose a new minimal bio-economic agent-based model designed to capture both macroscopic trends and individual farmer behavior. However, unlike existing models~\cite{paz_habitat_fragmentation, agrilove}, ours uniquely combines heuristic rules with aggregate data to approximate real-world scenarios. By employing relatively simple decision-making processes, our model successfully replicates historical trends in biodiversity loss, pesticide reduction, and farm size increases over the past 30 years, using French data. The novelty of our approach lies in its ability to strategically distill key factors influencing farmer decisions to replicate macro trends. This provides a simulation laboratory to investigate qualitatively the policy interventions that can mitigate farmland biodiversity decline. Indeed, the absence of  clear large-scale evidence of the effectiveness of usual agri-environmental policy instruments~\cite{PES_EU, PES_failure} advocates the need to conceive new models and policies to this end.}

{To conclude, the model and its results will be explained with the following structure.} In Section~\ref{sec:model}, we detail the model's structure and the rationale behind its design. In Section~\ref{sec:results}, we present the model's predictions under a ``business as usual" scenario, demonstrating that further declines in biodiversity occur due to continued farmland consolidation. Finally, we evaluate different policy options and show that the most effective approach for achieving both environmental and economic sustainability involves a combination of tighter pesticide regulations and support programs for small farms. In Section~\ref{sec:conclusion}, we conclude.

\section{A bio-economic agent-based model}\label{sec:model}

{In this section, we present the ecological and economic components of the model, along with the underlying data and its utilization in setting model parameters and initializing variables. The model description starts with an introduction adhering to the Overview, Design and Details (ODD) summary lines~\cite{ODD_grim_2020}, followed by the mathematical description of all submodels.\footnote{We opted for this approach instead of the full ODD+D framework, where +D stands for Decision-making~\cite{ODD}, because we believe that the simplicity of the models allows for complete transparency of the mathematical structure, which is crucial for a full understanding of the results.}}

\subsection{ODD Summary}

{The primary purpose of our model is twofold: (i) to understand the dynamics of agricultural landscapes in relation to biodiversity, and (ii) to suggest policy instruments aimed at supporting biodiversity conservation. Specifically, we address two questions: How do pesticide use and land consolidation influence biodiversity levels and ecological stability? Which policy instruments may be effective in curbing biodiversity degradation on a large scale? We ground our approach on real data by ensuring our model replicates historical trends in farm structure (using both average data and farm size distribution), pesticide application, and biodiversity indicators.}

{The model includes the following entities: farmers\footnote{Throughout this discussion, we will use the terms ``farms" and ``farmers" interchangeably.} and biodiversity.} Farmers produce a single standardized good, making decisions on two critical production factors: utilized agricultural land and pesticide use. We assume that farmers adjust their pesticide use to achieve specific yield targets, taking economic conditions into account. Profitability results from a combination of internal factors (pesticide efficiency, farm size) and external influences (market conditions, policies, and pest pressure). When profitable, farmers can invest in improving efficiency, potentially reducing pesticide use, and adjust their land holdings by renewing, terminating, or establishing new rental contracts\footnote{Note that land in the model is rented, in line with the fact that 80\% of utilized agricultural land in France is indeed rented~\cite{share_rent_UAA}}. 

{On the ecological side, a rescaled bird population index is used as an overall indicator for biodiversity. Indeed, bird population trends are often reliable indicators of environmental health~\cite{bird_as_good_index} as they occupy high trophic levels in food chains and reflect systemic changes~\cite{bird_and_agri}. In line with existing literature~\cite{Loreau_agri_ecosystem, Loreau_agri_ecosystem_3}, we characterize bird population evolution taking into account the degrading contribution of pesticide use~\cite{pesticide_biodiversity_2} and increased farm size~\cite{consolidation_biodiversity}. Increased farm size contributes to bird population decline via the reduction of field margins which are vital nesting sites. The biodiversity level determines potential pest damage which directly affects farmers' yield. Indeed, with greater biodiversity comes a lower risk of major pest outbreaks and the potential for higher crop yields~\cite{bird_and_pest_4, bird_and_pest, bird_and_pest_2, bird_and_pest_3}.}

{Regarding temporal resolution, a time step in the model represents one crop cycle (typically one year), and simulations cover the timeframe from 1990 to 2075\footnote{We select 2075 as a suitable temporal horizon,  distant enough to explore long-term effects given the model's inherent inertia, yet close enough for our approximations to remain reasonable throughout the period.}. The spatial resolution is at the national level, with no explicit spatial description;\footnote{{On the ecological side, this assumption can be partially justified by the fact that birds typically cover wider ranges compared to other species (such as insects or bees), making them less sensitive to region-specific details. On the economic side, we describe broader economic flows rather than localized spatial interactions to have a consistent macro level description in total land use and goods production. We acknowledge that this approach may simplify the complexities associated with biodiversity degradation and land consolidation. Therefore, we intend to explore ways of incorporating a spatial description in future versions of our model. However, considering that our model is intended as an exploratory tool for national-level dynamics, we believe that those assumptions do not significantly compromise the model's results.}} the scale refers to the number of farmers considered. The model's general characteristics are informed by the French agricultural sector, serving as a case study due to its significance in the European context. Economic aspects of farmers, considered constant such as production costs and economies of scale, are derived from field-crop businesses\footnote{Field-crop farming is the most extended land use in France by occupying about 30\% of the utilized agricultural area~\cite{FADN}}, whereas agronomic characteristics are modeled on wheat.\footnote{Wheat is the most cultivated cereal in France as it occupies 16\% of all agricultural land~\cite{FADN}} Free parameters have been calibrated to replicate historical trends for key variables, including the biodiversity index, average pesticide use, average farm size, farm size distribution, average yield, and market price.}

{Key processes in the model, repeated at each time step, include land consolidation, changes in pesticide use, and ecological responses (also illustrated in Fig.~\ref{fig:modelscheme}). Farmers adjust pesticide use and land holdings based on economic outcomes, while ecological components reflect these adjustments through changes in biodiversity indicators. Interactions among farmers emerge only in the land and goods markets, where the dynamics depend on collective actions.}

{The central design concept of the model is the integration of a microscopic representation of farmer decision-making with an aggregate biodiversity indicator. This approach allows biodiversity trends to emerge from agricultural practices and facilitates the development of effective policy instruments that may impact such trends.}

\subsection{Submodels}\label{sec:sub_mod}

\begin{figure}[t!]
\centering
\includegraphics[width=\textwidth]{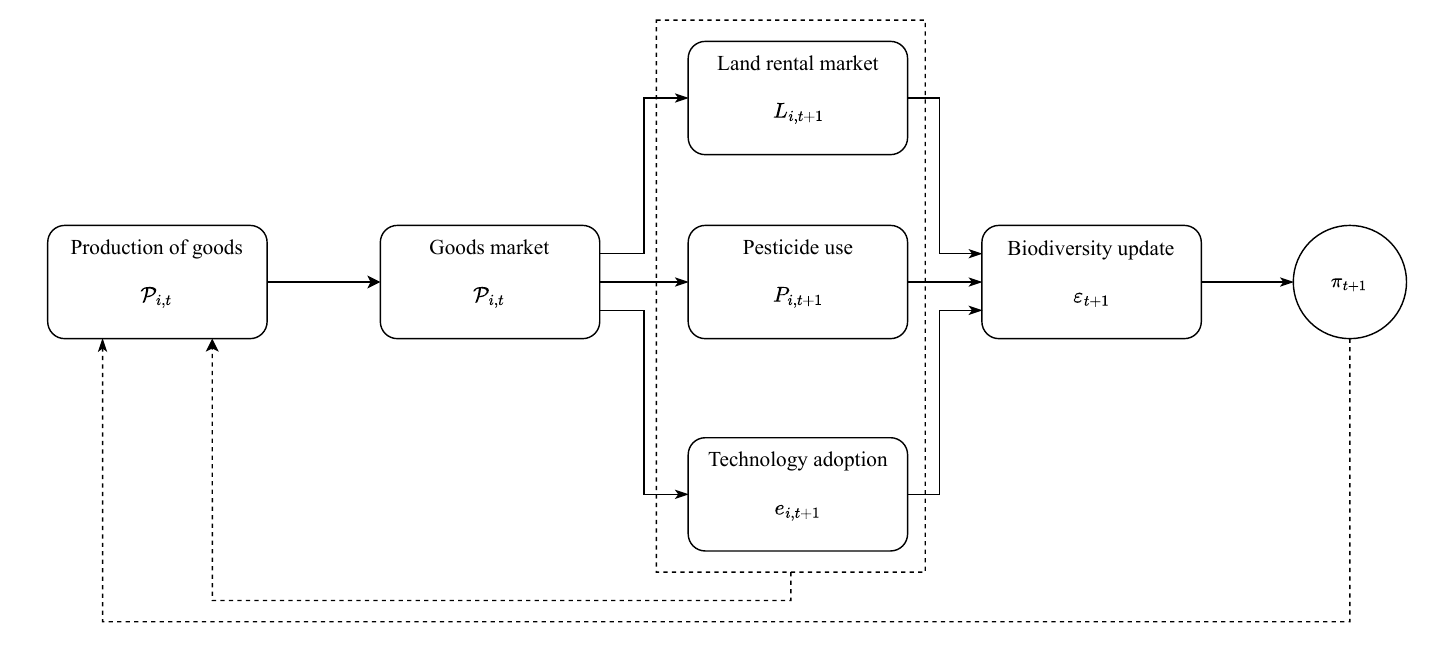}
\caption{Workflow of the model (see Section~\ref{sec:sub_mod}). Solid arrows represent direct influences, while dashed arrows indicate feedback loops affecting future production outcomes.}
\label{fig:modelscheme}
\end{figure}

{This section presents the model's components and their interactions, following the model's temporal sequence illustrated in Fig.~\ref{fig:modelscheme}. We outline the underlying hypotheses and provide the mathematical equations governing the dynamics.}

\subsubsection*{Production of goods}

The total production $Y_{i,t}$ of farmer $i$ at time $t$ is computed by multiplying his land area $L_{i,t}$ by his yield per hectare $y_{i,t}$:
\begin{eqnarray}
Y_{i,t} = L_{i,t} y_{i,t}.
\end{eqnarray}
The latter is assumed to be given by a  Mitcherliche-Baul production function~\cite{Mitcherliche_prod_func}:
\begin{eqnarray} \label{eq:prodfunc}
    y_{i,t} = y_{\text{max}} \left[1 - \pi_t \exp\left({ -\frac{e_{i,t}P_{i,t}}{P_{\mathrm{ref}}}}\right)\right](1 + \xi_{i,t}),
\end{eqnarray}
where $y_{\max}$ denotes the maximum attainable yield, according to agronomic choices not explicitly detailed here. $P_{i,t}$ denotes the pesticide application rate aimed at protecting the yield from pest damage $\pi_t$ (see Eq.~\ref{eq:pest}), while $e_{i,t}$ reflects the efficiency of pesticide use, with $P_{\mathrm{ref}}$ serving as a normalizing reference level. The efficiency $e_{i,t}$ encompasses variability in modern technology use (see below). Finally, $\xi_{i,t}$ is a white Gaussian noise accounting for further inter and intra-farmer variability. 

\subsubsection*{Goods market}
After harvest, farmers sell their grain at a central market where the price fluctuates according to supply and demand. The total grain supply $Y_{t}=\sum_i Y_{i, t}$ is the sum of the quantities produced by each individual farmer.\footnote{We shall henceforth systematically use the notation $A_t = \sum_i A_{i,t}$.} 
Demand $D$, on the other hand, is an external factor assumed constant in the model, consistent with the only weak fluctuations observed (over the past 30 years) and forecasted in Europe~\cite{future_food_demand}. In its simplest form we write 
\begin{eqnarray} \label{eq:price_update}
    p_{\mathrm{m},t} = p_{\mathrm{m},t-1} \left( 1 + \alpha \frac{D - Y_t}{D}\right)
\end{eqnarray}
for the price adjustment mechanism, where $\alpha$ reflects market frictions determining the speed at which the price adjusts to imbalances between supply and demand. Farmers' individual profit $\mathcal{P}_{i,t}$ is then given by
\begin{eqnarray}\label{eq:profit}
\mathcal{P}_{i,t} = p_{\mathrm{m},t} Y_{i,t} - \mathcal C_{i,t}  + \frac{L_{i,t}}{L_t} \mathcal S,
\end{eqnarray}
where production costs $\mathcal C_{i,t}$ write 
\begin{eqnarray}\label{eq:costs}
\mathcal C_{i,t} = L_{i,t} \left( p_{\mathrm{p}} P_{i,t} + \mathcal C_{\mathrm o} \right)  + (L_{i,t})^b \mathcal C_{\mathrm{no}},
\end{eqnarray}
and where  $\mathcal C_{\mathrm o}$ and $\mathcal C_{\mathrm{no}}$ are operational and non-operational expenses respectively. Operational costs encompass all expenses directly tied to producing grain per hectare, like fertilizer, seeds and fuel expenses (excluding pesticide costs). We assume these costs to be equal for all farmers. Pesticide application costs are factored in separately with $p_{\mathrm p} P_{i,t}$, where $p_{\mathrm p}$ denotes the price of  pesticide. 
Non-operational costs include fixed expenses such as land rental fees, depreciation, and wages. These costs generally scale sub-linearly with the amount of land ($
b<1$), reflecting economies of scale. Larger farms have lower average costs per unit because they can spread fixed costs over more units, benefit from bulk purchasing, and achieve other efficiencies~\cite{economy_of_scale, economy_of_scale_2}. Finally, the model incorporates direct payment of subsidies $\mathcal S$ proportional to the amount of land utilized by each farmer, in line with the Common Agricultural Policy (CAP) over the past three decades~\cite{CAP_no_price_support}.

\subsubsection*{Technology adoption}

Farmers strategically allocate a fraction $\eta$ of their profits to adopt advanced technologies that improve pesticide application efficiency. Examples of these technologies include drones for real-time monitoring of crop health, and predictive models from decision support systems to enhance pest control strategies~\cite{pesticide_efficiency}. {In our model, we assume that these technological advancements contribute to reduced pesticide use by minimizing overspraying, which subsequently benefits biodiversity. However, it is important to acknowledge that newer technologies could also lead to increased ecological harm. For instance, more powerful pesticides might target a wider range of species, potentially offering no net benefit for biodiversity. A discussion exploring this hypothesis is presented in Appendix~\ref{app:tech}.}

In line with the Evolutionary ABMs literature~\cite{tech_ABM}, we assume that the success of farmers' investments in technology adoption follows a Bernoulli trial, indicating that each attempt has a random chance of success. The probability of success $\mathbb{P}^{e}_{i,t}$ depends on the amount invested and the effectiveness in translating investment into successful innovations program ($P_{\mathrm{ref}}$) as
\begin{eqnarray} \label{eq:innov_prob}
\mathbb{P}^{e}_{i,t} = 1 - \exp\left(- \frac{ \eta \mathcal{P}_{i,t}}{\mathcal P_{\mathrm{ref}}}  \right).
\end{eqnarray}
If the investment of farmer $i$ at time $t$ is successful, the farmer's pesticide application efficiency level $e_{i,t}$ is increased by a random value $\upsilon_{i,t}$ as
\begin{eqnarray}
{e}_{i,t+1} = {e}_{i,t}+ \upsilon_{i,t},
\end{eqnarray}
where $\upsilon_{i,t}$ is drawn from a uniform distribution within a predefined range $[0, \upsilon_{\max}]$.

\subsubsection*{Land rental market}

Then, farmers strategically adjust their rented land based on their economic performance. The land rental market is described as a centralized market similar to that of Balman \emph{et al.}~\cite{land_market}. Farmers release and acquire land based on the difference between realized returns\footnote{Here the realized return on investment is computed as $r_{i,t} = \frac{(1 -\eta) \mathcal{P}_{i,t}}{\mathcal{C}_{i,t}}$.} and opportunity costs $r_{\mathrm{ref}}$, which correspond to the loss in returns of other investments that could have been made instead.

If actual returns  on investment $r_{i,t}$ are lower than $r_{\mathrm{ref}}$, farmers terminate some of their rental contracts to release the following quantity of land:
\begin{eqnarray}
L^{-}_{i,t} = \frac{\beta }{1+ \frac{r_{\mathrm{ref}}}{r_{\mathrm{ref}} - r_{i,t}} }L_{i,t},
\end{eqnarray}
where $\beta$ measures the maximum amount of released land per year. Farmer $i$'s available land for farming at the next period is then updated as
\begin{eqnarray}
 L_{i,t+1} =     L_{i,t} - L^-_{i,t}.
\end{eqnarray}
When the leftover land is less than $0.1$ hectares, the farmer terminates all rental contracts and leaves the market, thereby not  participating in the next year's production.

Conversely, if  opportunity costs are lower than  realized returns, the farmer is incentivized to sign new rental contracts. The amount of new prospective rented land is
\begin{eqnarray}
\tilde L^+_{i,t} =\frac{\beta }{1+ \frac{r_{\mathrm{ref}}}{ r_{i,t}-r_{\mathrm{ref}}} }L_{i,t} .
\end{eqnarray}
However, the actual new land acquisition should not exceed the total amount of land available for rent, computed as the sum of the leftover from the previous period and the land released by all agents 
\begin{eqnarray}
 L_{i,t+1} =     L_{i,t}  + L^+_{i,t}.
 \quad \textrm{where}\quad L^+_{i,t} = \tilde L^
+_{i,t} \min\left(1, \frac{\tilde L^+_{t}}{L^{\mathrm{leftover}}_{t-1}+L^{-}_{t}} \right).
\end{eqnarray}
Leftover land also needs to be updated as $ L^{\mathrm{leftover}}_{t}= L^{\mathrm{leftover}}_{t-1} + L^-_t- L^+_t$.

\subsubsection*{Pesticide use}

Remaining active farmers adjust their pesticide use according to past trends influencing their performance, rather than by strict economic optimization. Indeed, existing literature shows that pesticide adoption is largely influenced by a variety of behavioral factors~\cite{pesticide_app_behaviour}. {For simplicity, our model assumes that farmers update their pesticide use to meet an evolving yield target. This approach aligns with findings in agricultural economics that suggest farmers often prioritize specific yield targets as measures of success and viability, especially under uncertain conditions \cite{adaptation_farming_decision}.} Therefore, farmers update their pesticide use according to the following behavioral rule:
\begin{eqnarray} \label{eq:P_update}
    P_{i,t+1} = P_{i,t}\left(1+ \gamma \frac{\tilde y_{i,t}-y_{i,t}}{y_{i,t}}\right),
\end{eqnarray}
where  $\gamma$ is the speed of adjustment and  $\tilde y_{i,t}$ is the yield target (see below). {This approach takes into account the lack of complete information regarding both pest exposure levels and the efficiency of the technologies employed}. The yield target $\tilde y_{i,t}$ evolves according to market price fluctuations:
\begin{eqnarray}
\tilde y_{i,t+1} = \tilde y_{i,t} \left( 1 +  \lambda \frac{p_{\mathrm{m},t} - p_{\mathrm{m},t-1}}{p_{\mathrm{m},t}} \right).
\end{eqnarray}
This equation reflects the willingness of farmers to decrease or increase their production target according to market price: when the market price  has been decreasing (resp. increasing), farmers might lower (resp. increase) their yield target to reduce pesticide related costs. {Although this can sound like strong simplification of the farmers' decision-making process, it is commonly employed in ABMs as a plausible heuristic of production target setting by economic agents~\cite{mark0, DSK_macro_ABM}.}

\subsubsection*{Biodiversity update}
\label{subsec:ecolo}

{Finally, the actions of farmers influence biodiversity, described by the rescaled bird population $\varepsilon_t = M_t / M_0$, via a logistic equation in accordance with the existing literature~\cite{Mouysset2011, bird_logistic, Loreau_agri_ecosystem, Loreau_agri_ecosystem_3}:}

\begin{equation}
\label{eq:eps_dyn}
\varepsilon_{t+1} = \varepsilon_{t} + r_\varepsilon \left(1 -\frac{\varepsilon_{t}}{K_{t+1}} \right) \varepsilon_{t} ,
\end{equation}

{where $r_\varepsilon$ is the intrinsic growth rate of the bird population, and $K_t$ is a measure of the carrying capacity~\cite{dhondt1988carrying}. $K_t$ depends on the average pesticide use $\bar{P}_t$ and average farm size $\bar L_t$ as defined in the following equation:\footnote{Note that $\bar P_t$ is computed as a weighted average over farm sizes, in line with the usual reporting methods~\cite{FADN}.}}

\begin{equation}
\label{eq:car_cap}
K_{t+1}= \mu\frac{\bar{L}_0}{\bar{L}_{t+1}} + (1-\mu) \frac{\bar{P}_0}{\bar{P}_{t+1}},
\end{equation}

{where $\mu$ reflects their relative importance. The updated ecological conditions influence the pesticide-free typical yield loss due to pests $\pi_t$ via:}

\begin{equation}
\label{eq:pest}
\pi_{t+1} = \pi_0 \left( \frac{\varepsilon_0}{\varepsilon_{t+1}}\right)^a,
\end{equation}

{where the exponent $a$ determines the extent of biodiversity's impact on pest exposure and, consequently, on crop yields~\cite{yield_biodiversity} (see Appendix~\ref{app:bio_yield} for further discussion).}

\subsection{Real data and numerical simulations}

\subsubsection*{Data sourcing}

The data for calibrating and initializing the model are sourced from various databases. To characterize biodiversity, we use the common farmland bird index, which monitors the abundance of 34 farmland bird species across Europe~\cite{Bird_index_data}. Data on the agricultural sector's structure are derived from the Food and Agriculture Organization (FAO)'s ``Structural Data from Agricultural Censuses"~\cite{faostructuraldata}. This dataset is compiled through decennial monitoring of UN countries' agricultural censuses and includes detailed information on the size and number of agricultural holdings. Due to the lack of individual farmer data, agronomic information related to wheat production is sourced from the European Farm Accountancy Data Network (FADN)~\cite{FADN}. This provides average values for wheat yield and pesticide use. Finally, the wheat producer price index is used to track annual changes in selling prices for farmers. This index is constructed using price data in standardized local currency~\cite{FAO_price} (from FAO) on an annual basis.

\subsubsection*{Calibration} \label{subsubsec:calibration}

\begin{figure}[b!]
\centering
\includegraphics[width=\textwidth]{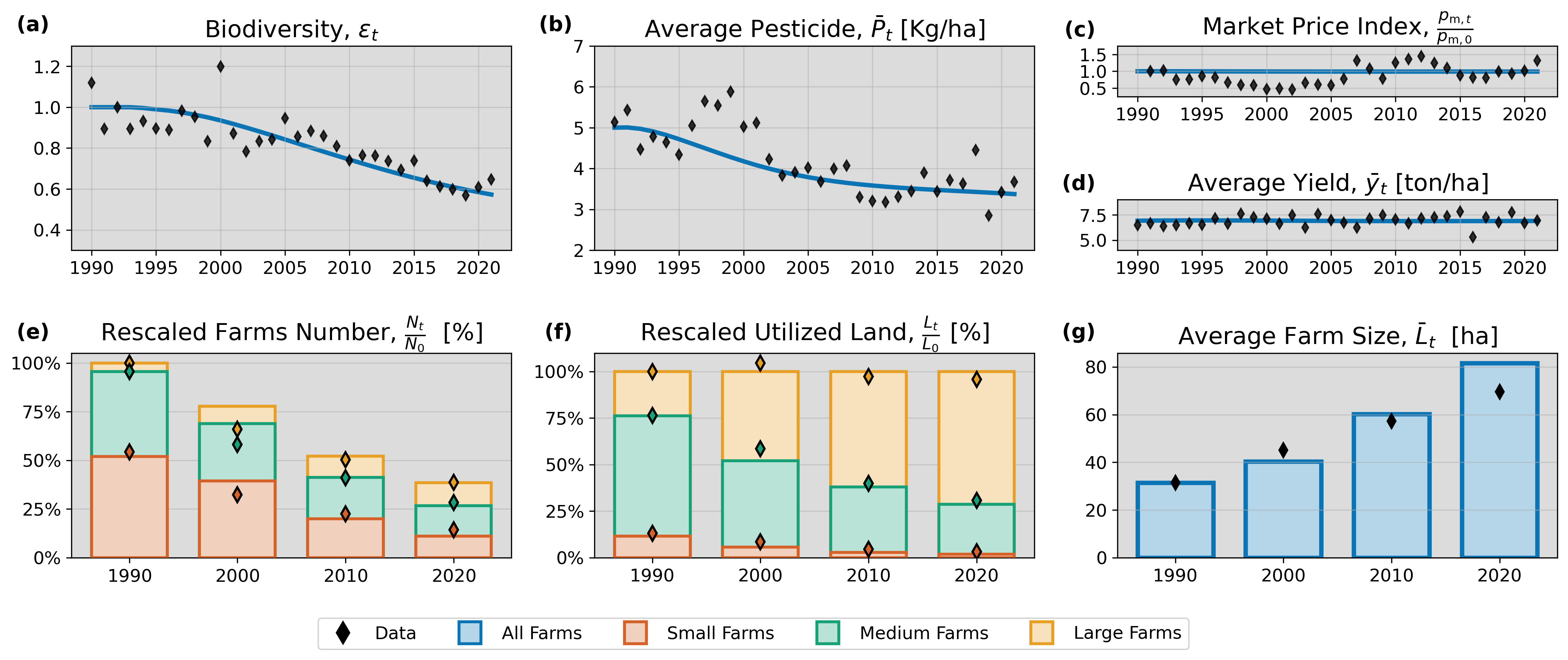}
\caption{Calibration of the model: diamonds represent real data, while blue lines and colored bars display the calibration results. (a) Biodiversity degradation (EU common farmland bird index~\cite{Bird_index_data}). (b) Decline in average pesticide use (FADN~\cite{FADN}). (c) Market price index (FAO~\cite{FAO_price}). (d) Average yield (FADN~\cite{FADN}). (e-g) Land consolidation illustrated with FAO data~\cite{faostructuraldata} in terms of (e)  total number of farmers, (f) total utilized land for different farm sizes (small farms: less than 20 ha, medium farms: 20-100 ha, large farms: more than 100 ha), and (g) average farm size.}
\label{fig:calibration}
\end{figure}

Model parameters are categorized into three groups (see Table~\ref{tab:params}): (i) those directly inferred from data and literature, henceforth called \textit{measured}  parameters (\textcolor{dataColor}{M}), (ii) those set by hand to ensure consistent dynamics, namely \textit{estimated}  parameters (\textcolor{estimatedColor}{E}), and (iii) those obtained through calibration to reproduce trends during the 1990-2021 period, or \textit{calibrated} parameters (\textcolor{calibratedColor}{C})  (see Fig.~\ref{fig:calibration}). The main trends within the agricultural landscape are biodiversity degradation (Fig.~\ref{fig:calibration}a), decline in average pesticide use due to technological innovation depicted (Fig.~\ref{fig:calibration}b), and land consolidation (Figs.~\ref{fig:calibration}e-g).

Although derived from existing data, the values of the parameters in the \textcolor{dataColor}{M} category are influenced by model assumptions and simplifications. For example, some quantities—such as non-operating costs, operating costs, and pesticide prices—fluctuate over time, but are assumed constant and equal to representative values over the 1990-2021 period.

Parameters in the  \textcolor{estimatedColor}{E} category are set to reasonable values that ensure consistency between the data and the model's functional forms.  For example, the ecological parameter $\mu$ in Eq.~\ref{eq:car_cap} is set to reproduce the biodiversity curve using Eq.~\ref{eq:eps_dyn} with historical values of average pesticide use and farm size.\footnote{Given the noisy nature of the data, we use a rolling average as a more stable proxy of these trends.} Similarly, the value of $a$ (Eq.~\ref{eq:pest}) was derived using biodiversity interpolated values, assuming a pest exposure value of 0.3 in 1990 and 0.4 in 2021~\cite{potential_pest_damage_report_EU}. On the economic side, $P_{\mathrm{ref}}$ is computed by inverting Eq.~\ref{eq:prodfunc} using average values from 1990 data and $\bar e_0=1$, while $r_{\mathrm{ref}}$ and $D$ are based on the equilibrium hypothesis (see \textit{Initialization} below).

Parameters in the \textcolor{calibratedColor}{C} category are primarily related to technology adoption and behavioral aspects of the model. Realistic values for these parameters are not directly available due to the high-level description of the model, and they cannot be estimated as they depend intrinsically on the model's dynamic realization. 
Therefore, we opted for calibrating these parameters to ensure the model replicates well the major trends observed in the reference data from 1990 to 2021. {The calibration  was done through least-squares minimization, with additional methodological details and sensitivity analysis provided in the Appendix~\ref{app:calibration_sensitivity}. The results are given in Table\ref{tab:params} and Figure~\ref{fig:calibration}.}

\begin{table}[t!]
    \centering
    \begin{tabular}{lcllcc}
        \multicolumn{6}{c}{~} \\ \hline
        Section & Notation & Description & Value & Group & Source \\ \hline
        Initialization & $N_0$ & Number of farms & {300000} & \textcolor{dataColor}{M} & \cite{faostructuraldata} \\ 
        & {$L_0$} & Total agricultural land \footnotesize{$[\mathrm{ha}]$} & {10e6} & \textcolor{dataColor}{M} & \cite{faostructuraldata} \\ 
        & $\bar P_0$ & Average pesticide use \footnotesize{$[\mathrm{kg \cdot ha}^{-1} \mathrm{year}^{-1}]$}  & 5 & \textcolor{dataColor}{M} & \cite{FADN} \\
        & $\bar y_0$ & Average yield \footnotesize{$[\mathrm{ton \cdot ha}^{-1} \mathrm{year}^{-1}]$} & 7 & \textcolor{dataColor}{M} & \cite{FADN} \\
         & $\pi_0$ & Potential pest damage \footnotesize{$[ \mathrm{year}^{-1}]$} & 0.3 & \textcolor{dataColor}{M} & \cite{potential_pest_damage_report_EU} \\  
        ~ & $r_{0}$ & Average return on investment \footnotesize{$[ \mathrm{year}^{-1}]$} & 0.05 & \textcolor{dataColor}{M} & \cite{FADN} \\
        ~ & $D$ & Demand \footnotesize{$[\mathrm{ton} \cdot \mathrm{year}^{-1}]$} & {7e7} & \textcolor{estimatedColor}{E} & \\
        \hdashline[.4pt/1pt]
        Ecological & $r_\varepsilon$ & Intrinsic growth rate \footnotesize{$[ \mathrm{year}^{-1}]$} & 0.1 & \textcolor{dataColor}{M} & \cite{bird_logistic} \\ 
        factors & $\mu$ & Carrying capacity weights  & 0.9 & \textcolor{estimatedColor}{E} & ~ \\
        ~ & $a$ & Pest-biodiversity nonlinearity & 0.5 & \textcolor{estimatedColor}{E} & ~ \\
        \hdashline[.4pt/1pt]
        Production & $y_{\max}$ & Maximum yield \footnotesize{$[\mathrm{ton \cdot ha}^{-1} \mathrm{year}^{-1}]$} & 8.5 & \textcolor{dataColor}{M} & \cite{maximum_yield} \\ 
        factors & $P_{\mathrm{ref}}$ & Reference pesticide \footnotesize{$[\mathrm{kg \cdot ha}^{-1} \mathrm{year}^{-1}]$} & 10 & \textcolor{estimatedColor}{E} & ~ \\ 
        ~ & std($\xi$) & Production noise  & 0.05 & \textcolor{dataColor}{M} & \cite{yield_noise} \\
        \hdashline[.4pt/1pt]
        Other economic & $\alpha$ & Price frictions & 0.08 & \textcolor{calibratedColor}{C} & ~ \\ 
        factors & $p_{\mathrm p}$ & Pesticide price \footnotesize{$[\text{\euro} \cdot \mathrm{kg}^{-1}$]} & 10 & \textcolor{dataColor}{M} & \cite{pesticide_price} \\ 
        ~ & $\mathcal{C}_{\mathrm O}$ & Operating costs \footnotesize{$[\text{\euro} \cdot \mathrm{ha}^{-1}\mathrm{year}^{-1}$]} & 500 & \textcolor{dataColor}{M} & \cite{EU_Cereal_costs} \\ 
        ~ & $\mathcal{C}_{\mathrm{NO}}$ & Non-operating costs \footnotesize{$[\text{\euro} \cdot \mathrm{ha}^{-1}\mathrm{year}^{-1}$]} & 600 & \textcolor{dataColor}{M} & \cite{EU_Cereal_costs} \\ 
        ~ & $b$ & Economy of scale & 0.9 & \textcolor{dataColor}{M} & \cite{economy_of_scale} \\ 
        ~ & $\mathcal{S}$ & Total subsidies budget \footnotesize{$[\text{\euro} \cdot \mathrm{year}^{-1}$]} & {5e9} & \textcolor{dataColor}{M} & \cite{CAP_history} \\ 
        \hdashline[.4pt/1pt]
        Technology & $\eta$ & Profit share for technology & 0.15 & \textcolor{calibratedColor}{C} & ~ \\ 
        adoption & \footnotesize{$\mathcal{P}_{\mathrm{ref}}$} & Reference profit for technology \footnotesize{$[\text{\euro} \cdot \mathrm{year}^{-1}]$} & 1000 & \textcolor{calibratedColor}{C} & ~ \\ 
        ~ & $\upsilon_{\max}$ & Maximum efficiency gain \footnotesize{$[ \mathrm{year}^{-1}]$}  & 0.1 & \textcolor{calibratedColor}{C} & ~ \\ 
        \hdashline[.4pt/1pt]
        Behavioural & $\beta$ & Land adjustment speed  \footnotesize{$[ \mathrm{year}^{-1}]$} & 0.45 & \textcolor{calibratedColor}{C} & ~ \\ 
        factors & \footnotesize{$r_{\mathrm{ref}}$} & Reference return on investment \footnotesize{$[ \mathrm{year}^{-1}]$} & 0.05 & \textcolor{estimatedColor}{E} & \\ 
        ~ & $\gamma$ & Pesticide adjustment speed & 2 & \textcolor{calibratedColor}{C} & ~ \\ 
        ~ & $\lambda$ & Yield target adjustment speed & 0.2 & \textcolor{calibratedColor}{C} & ~ \\ \hline
    \end{tabular}
    \caption{Parameters of the model. We indicate their units, value, estimation method, and source (see Section~\ref{subsubsec:calibration}).}
    \label{tab:params}
\end{table}

\subsubsection*{Initialization} 

Having set the parameters, the model can now be initialized. On the economic side, {the initial number of farmers is determined to approximate the number engaged in fieldcrop farming in 1990, see Table~\ref{tab:params}}.  The distribution of their land holdings is derived from from the FAO's "Structural Data from Agricultural Censuses" dataset for 1990. For agronomic information related to wheat production, we use average values from the FADN dataset. Each farmer is assigned an initial pesticide use and yield as follows:
\begin{eqnarray}
    \begin{array}{lr}
    P_{i, 0} = \bar P_{0} (1 + \zeta_i)\\
    y_{i, 0} = \bar y_{0} (1 + \psi_i)
  \end{array}
  \quad \quad \text{where} \quad \quad \zeta_i, \psi_i \in \mathcal{N}(0, 0.1)
\end{eqnarray}
with $\mathcal{N}(0, 0.1)$  a Gaussian distribution with zero mean and a standard deviation of 0.1. Further assuming an initial pest exposure of 30\% (see above)~\cite{potential_pest_damage_report_EU}, we calculate the initial efficiency rate and total costs using Eqs~\ref{eq:prodfunc} and~\ref{eq:costs}, respectively. The initial market price is set to reflect an average return on investment equal to the opportunity cost, ensuring a competitive agricultural sector at the beginning of the period. 
Finally, the initial demand is set equal to the total initial production, simulating a market close to equilibrium.

\subsubsection*{Simulation}

With all parameters set and variables initialised, various scenarios can be explored through extensive numerical simulations. For each scenario, we conduct 10 simulations with different random seeds. The results presented are Monte Carlo averages, with their relative standard errors omitted due to their negligible values (below the line width). It is important to note that we do not account for global fluctuations in either ecological or economic aspects, so the relative differences between simulations are solely due to farmer-to-farmer variability. Given the large number of farmers, this variability averages out when examining macro trends.

Instead of relying on a single indicator, we shall propose a stratified analysis to examine each variable's role through its interactions with others, providing a more comprehensive understanding of the dynamics of the model.

\section{Results} \label{sec:results}

We begin by examining a baseline scenario without additional policy interventions. Our model predicts a decline in biodiversity by 2075 of 10\% compared to 2021, and a concerning 55\% decline compared to 1990 levels. {While these values should be considered as approximate given the nature of our analysis, the significant drop underscores the urgent need for policy changes to protect and, when possible, restore biodiversity.} To address this challenge, we explore two potential policy options: (i) reducing pesticide use and (ii) implementing subsidies for small farmers. Our analysis shows that both approaches positively impact overall biodiversity, albeit with some caveats. The most effective strategy appears to be a combined approach that leverages the strengths of both policies.

\begin{figure}[t!]
\centering
\includegraphics[width=\textwidth]{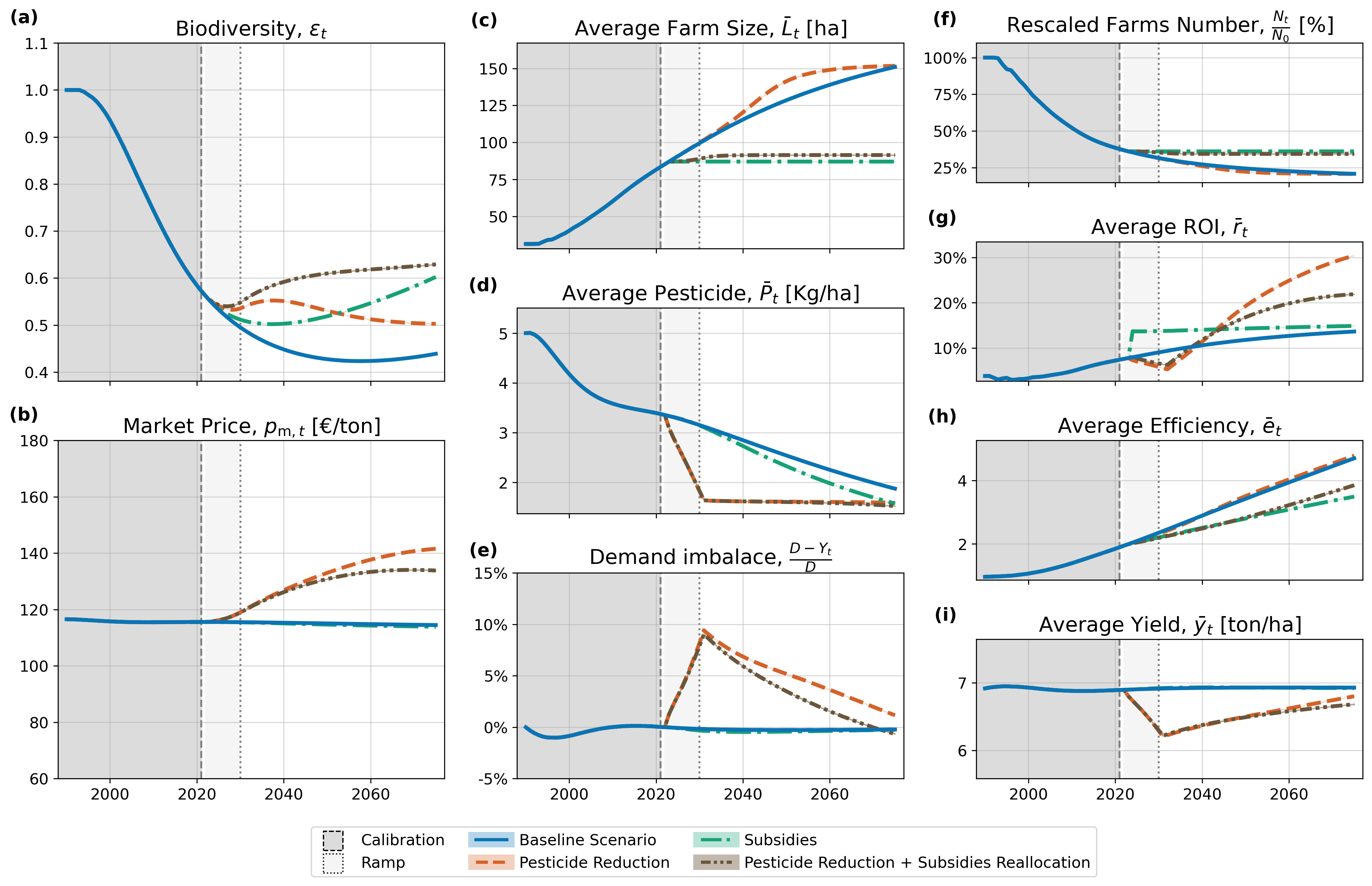}
\caption{Numerical simulations. The gray-shaded area indicates the calibration period (1990-2021).  In the baseline scenario (solid blue lines), biodiversity continues to decline due to intensified land consolidation. The pesticide reduction scenario (dashed orange line) shows initial biodiversity recovery, followed by a decline caused by intensified land consolidation due to reduced yields. The light-gray-shaded area marks the 2022-2030 ramp to halving pesticide.
The subsidy program favoring small farmers (dash-dotted green line) effectively halts land consolidation, leading to improved biodiversity outcomes. The combined approach of pesticide reduction with subsidies for small farmers (dash-dot-dotted brown line) appears as the most promising solution for biodiversity conservation by mitigating the negative effects of land consolidation while benefiting from pesticide reduction.}
\label{fig:scenarios}
\end{figure}

\subsection{Baseline scenario}

The business-as-usual predictions of our model are represented by the solid blue lines in Fig.~\ref{fig:scenarios}.
In this baseline scenario, the ongoing biodiversity degradation process (Fig.~\ref{fig:scenarios}a) proceeds in two distinct phases. In the short term (2022-2050), biodiversity drops to 42\% of its 1990 levels. However, in the latter part of the simulation (2050-2075), we observe a slight recovery to eventually reach 45\%.

The initial decline is primarily driven by intensified land consolidation. Although  pesticide use (Fig.~\ref{fig:scenarios}d)  decays over time due to increased efficiency (Fig.~\ref{fig:scenarios}h), the positive effect of such decay on biodiversity is offset by the increase of average farm size (Fig.~\ref{fig:scenarios}c). 
 The land consolidation, where the number of farms continuously decreases (Fig.~\ref{fig:scenarios}f) with smaller farms being absorbed by larger ones, is driven by the following mechanism. Since the market price remains relatively constant (Fig.~\ref{fig:scenarios}b), farmers must compensate for the increase in pest exposure, due to biodiversity loss~(Eq.~\ref{eq:pest}), by enhancing efficiency to maintain their marginal profit. However, investments in technology adoption depend on current profitability~(Eq.~\ref{eq:innov_prob}), favoring larger farmers with greater margins due to economies of scale. Over time, small struggling farms terminate rental contracts, sacrificing further economies of scale and profitability, leading to sector abandonment and further biodiversity decline. 

In the long term, the number of farmers stabilizes as most farms reach a size sufficient to secure a return on investment greater than the opportunity cost (Fig.~\ref{fig:scenarios}g). Consequently, in this phase, the continued reduction in pesticide use does explain the slight recovery of biodiversity. {While the exact 
timing of this stabilization should not be taken to seriously, it seems reasonable when considering that European farmers face international competition from countries where the average farm size (and thus their capacity to exploit economies of scale) reaches hundreds of hectares~\cite{faostructuraldata}. Therefore, this market pressure on small farmers, which has already driven land consolidation and ecological degradation over the last 30 years, may represent a serious threat to biodiversity in the coming decades~\cite{consolidation_biodiversity}.}

Lastly, it is noteworthy that, as mentioned above, the market price remains nearly unchanged because total production remains aligned with demand (Fig.~\ref{fig:scenarios}e). This equilibrium is achieved by maintaining yield levels (Fig.~\ref{fig:scenarios}i) equivalent to the period 1990-2021, indicating that average efficiency gains offset the effects of  increased pest exposure. 

\subsection{Pesticide reduction}

Although the decay rate is indeed slowing down, the forecasted biodiversity decline in the baseline scenario remains unacceptable. 
Indeed, biodiversity is critical for ecosystem functioning and supporting essential services such as water provision and air purification, vital for societal well-being~\cite{bio_ecosystem_wellbeing, bio_ecosystem_wellbeing_2}, though these aspects are not the focus of this study.
One widely discussed policy proposal to enhance biodiversity is reducing pesticide usage~\cite{Farm_to_Fork,pesticide_tax,organic_EU}. Various methods can be thought of, such as pesticide taxes~\cite{pesticide_tax} or incentives for organic farming~\cite{organic_EU}. Our analysis focuses on a straightforward approach: mandating a gradual reduction in pesticide use by farmers. Specifically, we model a policy requiring farmers to halve their pesticide usage linearly  over 2022 to 2030 the period,\footnote{This is achieved via the following equation: $P_{i, t+1} = \min \left\{ P_{i,2021} \min\left[\max\left(\frac{t+1-2021}{2030-2021}, 0\right), 1\right], P_{i,t}\left(1+ \gamma \frac{\tilde y_{i,t}-y_{i,t}}{y_{i,t}}\right) \right\} $.} in line with the European Union considerations~\cite{Farm_to_Fork}.

The outcomes of the pesticide reduction policy are illustrated as orange dashed lines in Fig.~\ref{fig:scenarios}. In the beginning, biodiversity recovers rapidly (Fig.~\ref{fig:scenarios}a) due to the immediate effect of reduced pesticide use (Fig.~\ref{fig:scenarios}d). However, this initial boost is quickly followed by a strong decline. Indeed, reduced pesticide use leads to an estimated 10\% decrease in crop yield (Fig.~\ref{fig:scenarios}i), consistent with existing literature~\cite{yield_reduction_by_pesticide_reduction_EU_report}. Consequently, the reduced yield results in a significant drop in return on investment for farms of all sizes (Fig.~\ref{fig:scenarios}g). Smaller farms, already operating with narrow profit margins, are particularly affected and tend to exit the farming sector even faster. This accelerates land consolidation, as evidenced by a decrease in the number of farmers (Fig.~\ref{fig:scenarios}f) and an increase in average farm size compared to the baseline scenario (Fig.~\ref{fig:scenarios}c). 
Ultimately, exacerbated land consolidation neutralizes the positive impacts of pesticide reduction, resulting in a net decrease in biodiversity that falls below 2021 levels.

Finally, we observe an increase in market price  which can be attributed to an underproduction regime (Fig.~\ref{fig:scenarios}e) resulting from decreased yields (Fig.~\ref{fig:scenarios}i) without corresponding expansion in cultivated land. However, and perhaps surprisingly, such increase in price is not sufficient to outweigh the profit degradation due to yield loss, resulting on average in a lower ROI (Fig.~\ref{fig:scenarios}g). This can be explained by the influence of international competition, which is factored in here as friction in the price dynamics ($\alpha < 1$ in Eq.~\ref{eq:price_update}). Indeed, in a global market, one expects that local decreases in crop yield are offset by higher production levels in other countries not subject to the same pesticide reduction policies. 

{While the magnitude of the indirect economic pressure on small farmers exerted by pesticide reduction policies may vary depending on region-specific properties, we believe these effects can be significant. This highlights the need for more integrated model and policy frameworks~\cite{pesticide_reduction_policy, EU_biodiversity_plan} that consider the interplay between pesticide use, land consolidation, and the economic viability of small farming systems.}

\subsection{Subsidies}

As argued above, land consolidation significantly influences biodiversity dynamics. Therefore, it is logical to consider a policy involving subsidies that favors small farmers to maintain a diverse landscape. To this end, we introduce a subsidy program in the form of a direct payment to all farmers (in addition to the existing subsidies per ha), providing a 200-euro coupon regardless of farm size~\footnote{{This modifies Eq.~\ref{eq:profit} by introducing a constant term $\mathcal{S}_{\mathrm{P}} =200\text{\euro}$.}}. This policy inherently benefits small farmers due to the marginal increase in profit it provides.

The results of this policy are shown in Fig.~\ref{fig:scenarios} as green dash-dotted lines. The number of farmers (Fig.~\ref{fig:scenarios}f) rapidly stabilizes after the application of the new policy, as does the average farm size (Fig.~\ref{fig:scenarios}c). Consequently,  biodiversity degradation slows down compared to the baseline scenario to eventually increase  (Fig.~\ref{fig:scenarios}a). In addition, biodiversity restoration positively affects pesticide use. As pest exposure decreases accordingly to Eq.~\ref{eq:pest}, the amount of pesticide required to achieve the same yield also decreases (Figs.~\ref{fig:scenarios}d and i). Finally, the reduction in pesticide use is reflected in slightly lower prices (Fig.~\ref{fig:scenarios}b), due to the slow adjustment behavior of farmers, which results in a phase of overproduction.

The results of this section highlight the critical role of favoring small farmers in halting land consolidation and enhancing biodiversity.

\subsection{Combined policy: pesticide reduction and subsidies reallocation}

Thus far, we have demonstrated that biodiversity decreases under a business-as-usual scenario. The pesticide policy alone is ineffective in improving the situation due to its associated impact on land consolidation. Conversely, a subsidy policy favoring small farmers tends to mitigate land consolidation.

\begin{figure}[t!]
\centering
\includegraphics[width=\textwidth]{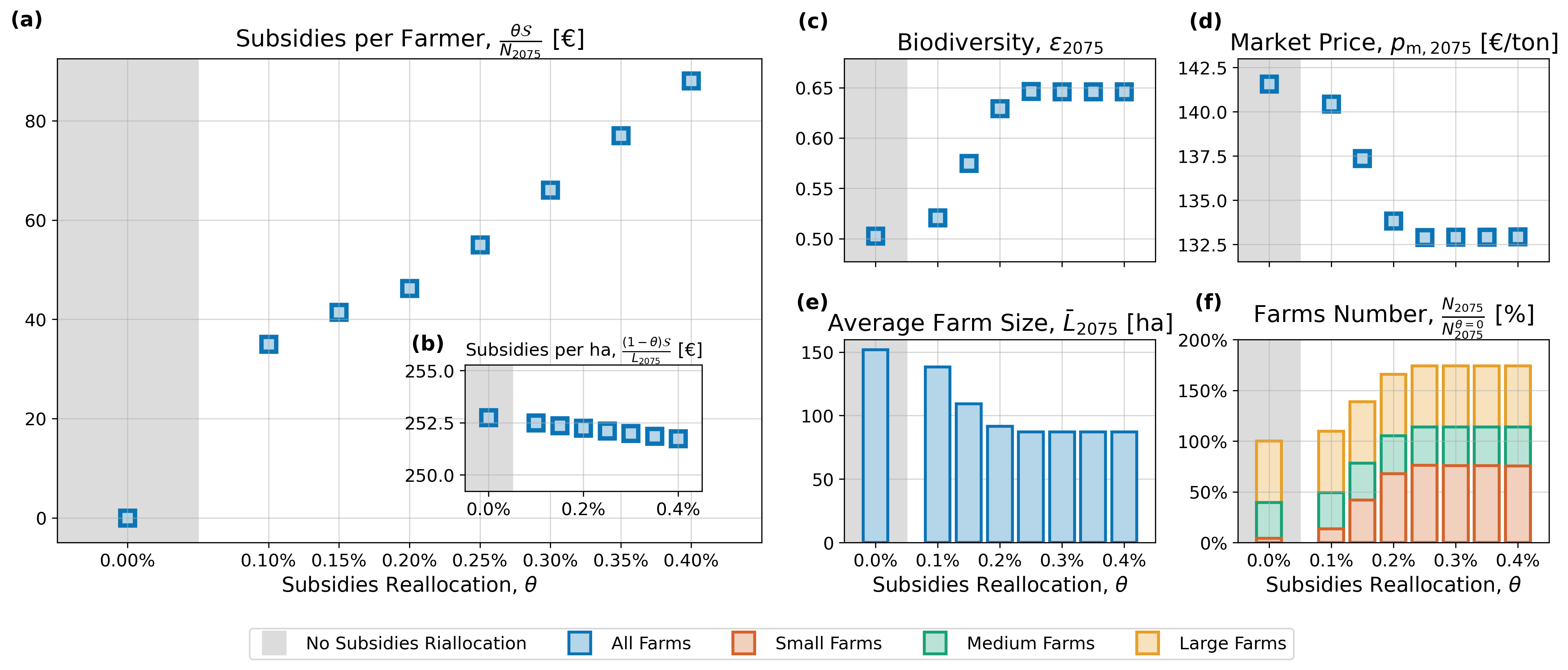}
\caption{Impact of the combined policy in 2075 for different levels of subsidies reallocation. (a) Subsidies per farmer as function of the percentage of subsidies reallocation. (b) Subsidies per hectare negligible decrease. (c) Biodiversity improves significantly with increased subsidies reallocation as  the average farm size (e) decreases, indicating a reduction in land consolidation. However a plateau is reached when the numbers of farmers (f) is no longer affected by a further increase in subsidies. (d) The market price decreases slightly with higher subsidies reallocation, as a consequence of biodiversity recovery, ultimately benefiting societal well-being. }
\label{fig:subsidies_analysis}
\end{figure}

Regarding policy makers motivations and constraints, on one hand, they may seek to reduce pesticide use regardless of biodiversity benefits given the well-documented negative impact of pesticide use on human health~\cite{pesticide_human_health}. On the other hand, they may be reluctant to add new subsidies to the already existing ones due to funding constraints. Therefore, we {explore the consequences of} a coordinated policy that targets (i) pesticide reduction (through the same mechanism as before), while (ii) reallocating a fraction of existing subsidies from a per-hectare basis to a per-farmer basis \footnote{{This is achieved by modifying Eq~\ref{eq:profit} as $\mathcal{P}_{i,t} = p_{\mathrm{m},t} Y_{i,t} - \mathcal C_{i,t}  + \frac{L_{i,t}}{L_t} (1-\theta)\mathcal S +  \frac{1}{N_t}\theta\mathcal S$.}}. The results of this combined policy, assuming a reallocation of 0.3\% of existing subsidies, are reported in Fig.~\ref{fig:scenarios} (dash-dot-dotted brown line). This reallocation choice corresponds to a new subsidy of 65 euros per farmer and a negligible different in subsidy per hectare (see~Figs~\ref{fig:subsidies_analysis}a-b).

The results of this combined policy, assuming a reallocation of 0.3\% of existing subsidies, are reported in Fig.~\ref{fig:scenarios} (dash-dot-dotted brown line). This reallocation choice corresponds to a new subsidy of 65 euros per farmer and a negligible difference in subsidy per hectare (see Figs~\ref{fig:subsidies_analysis}a-b).

The positive impacts of both policies pesticide reduction and targeted subsidies are successfully combined. Biodiversity rapidly increases due to pesticide reduction (Fig.~\ref{fig:scenarios}a), and, unlike with the pesticide reduction policy alone, such  increase is stabilized by the action of the subsidies. Indeed, the number of farmers (Fig.~\ref{fig:scenarios}f) and the average farm size (Fig.~\ref{fig:scenarios}c) stabilize. We also observe that market prices benefit from this increase in biodiversity (Fig.~\ref{fig:scenarios}b). Indeed, due to biodiversity recovery, lower pesticide use is required to achieve the same yield targets (Eq.~\ref{eq:prodfunc}), reducing the duration and magnitude of the underproduction phase (Fig.~\ref{fig:scenarios}e). This leads to an overall price level lower than under the pesticide reduction policy alone, thereby benefiting society as a whole.

As stated above, the results of Fig.~\ref{fig:scenarios} are given for a reallocation fraction of 0.3\% of pre-existing subsidies. One may wonder how varying this reallocation fraction affects the overall outcomes. In Fig.~\ref{fig:subsidies_analysis}, we report  results at the end of the period (year 2075) for different reallocation fractions. When the reallocation fraction is too small, the positive effects are limited, with biodiversity showing only a slight improvement over the pesticide reduction policy alone (Fig.~\ref{fig:subsidies_analysis}c). However, increasing this fraction by a few basis points leads to a strong increase in overall benefits, as highlighted by higher levels of biodiversity and lower prices (Fig.~\ref{fig:subsidies_analysis}d). Increasing further the reallocation fractions is not necessary, indeed benefits saturate as the distribution of farm sizes is no longer affected (Figs~\ref{fig:subsidies_analysis}e-f).\footnote{Recall that we have not allowed for entry of new farmers in this version of the model.}

{While the evaluation of the effects of the subsidy reallocation may not be quantitatively accurate and requires more robust exploration, we believe that the identification of possible feedback loops from biodiversity recovery to societal well-being, as described here, is crucial in halting the ongoing environmental degradation. Indeed, as advocated by an increasing number of researchers, the environmental challenges we face require an analysis that goes beyond first-order economic effects~\cite{bio_policy_beyond_economic}. To this end, minimal models like ours are useful tools for establishing a qualitative phenomenology of such complex phenomena by identifying basic mechanisms and potentially important, yet hidden, positive feedback loops.}

\section{Discussion}\label{sec:conclusion}

Using the French agricultural landscape as a case study, we present a parsimonious, integrated bio-economic model based on historical data. Our approach stands out by leveraging micro-foundations to understand past and future trends in biodiversity loss, pesticide use, and land consolidation. {Moreover, the model is constructed to evaluate policy interventions, analyzing their potential effects on ecological and economic sustainability}.

Our framework specifically incorporates feedback loops between bird population dynamics (as a proxy for overall biodiversity) and a granular description of farmer decision-making processes. The model projects a significant decline in biodiversity under a business-as-usual scenario, primarily driven by land consolidation. This highlights the urgent need for effective policy measures, given the well-documented role of biodiversity in societal well-being. To address this challenge, we analyze the potential impact of two policy interventions: pesticide reduction and subsidies {favoring small} farmers. Both policies demonstrate their capacity to positively impact biodiversity, although through different mechanisms.

Pesticide reduction initially boosts biodiversity but ultimately leads to decreased crop yields and financial stress on small farms, accelerating land consolidation and negating initial biodiversity gains. Consequently, the most effective approach appears to be a combined strategy that incorporates both pesticide reduction and targeted subsidies for small farmers, which can be achieved by reallocating a modest fraction of the already existing subsidies.  By mitigating the economic pressures on small farms, these subsidies can help maintain a more diverse agricultural landscape, thereby balancing the benefits of pesticide reduction with economic stability for small farms \footnote{The idea that the most effective policies consist in combining policy instruments that are complementary (e.g. duo of subsidies and restrictions) was recently discussed in~\cite{stechemesser2024climate}.}.

{While our minimal model provides valuable insights into policy interventions by strategically focusing on key factors that influence biodiversity and farmer decisions, it is essential to note that our results should not be interpreted as precise quantitative forecasts for real-world scenarios. Consequently, the policy recommendations drawn from this model should be corroborated by addressing the current model limitations. We identify three main directions to improve the model's applicability and robustness: (i) including diverse farming types, (ii) refining the representation of farmer decision-making processes, and (iii) incorporating a spatial dimension.}

{Indeed, it will be important to broaden the model's scope to encompass the diverse range of prevalent farming practices.  While a significant number of farmers primarily concentrate on field crops, other farming types, such as permanent crops and grazing livestock, are also significant within the French agricultural landscape. Each of these farming systems involves distinct farmer decision-making processes and has unique impacts on biodiversity~\cite{different_farm_systems, different_farm_systems_2}.}

{Accurately capturing the intricate nature of real-world agricultural decision-making will require refining the model's representation of farmer behavior. Existing literature highlights the crucial role of integrating behavioral factors and socio-economic influences into realistic representation  of farmer decision-making processes~\cite{review_ABM_farming_decision, adaptation_farming_decision}. Indeed, factors such as risk aversion, market trends, and evolving policy frameworks are expected to impact farmer choices~\cite{risk_farming_decision, sustainable_farming_decision}.}

{Lastly, incorporating a spatial dimension will enhance the model's ability to reflect the geographical heterogeneity of agricultural landscapes. These landscapes are characterized by a mosaic of land use types, each influencing biodiversity through features like hedgerows, field margins, and natural habitats~\cite{semi_natural_biodiversity}.  A spatially explicit model would be better equipped to account for these characteristics and their influence on the spatio-temporal dynamics of biodiversity and economic interactions, thereby facilitating more detailed and spatially-informed policy assessments.}

{In conclusion, while our current model serves as a valuable tool for policy exploration, addressing the identified areas for extension is essential to enhance its effectiveness as a policy support tool in real-world settings. Future models that incorporate these features will most likely achieve greater applicability and offer region-specific policy recommendations tailored to the unique challenges faced by diverse agricultural landscapes.}

\section*{Acknowledgments }

We deeply thank Karl Naumann-Woleske who contributed to the early stages of this work, as well as Jean-Philippe Bouchaud and Michel Loreau for their insightful suggestions.  We are also grateful to Damien Challet, Alexandre Darmon, Pierre Lenders and Antoine Mandel for fruitful discussions.
This research was conducted within the Econophysics \& Complex Systems Research Chair, under the aegis of the Fondation du Risque, the Fondation de l’École polytechnique, the École polytechnique and Capital Fund Management.

\newpage

\bibliographystyle{unsrt}
\bibliography{new_bib}

\begin{thebibliography}{100}

\bibitem{ipbes_report_bio_decline}
IPBES.
\newblock {Summary for policymakers of the global assessment report on biodiversity and ecosystem services}, November 2019.

\bibitem{Species_biomass_decline}
R.~E.~A. Almond, M.~Grooten, and T.~Petersen.
\newblock Living planet report 2020 - bending the curve of biodiversity loss.
\newblock Report, WWF, Gland, Switzerland, 2020.

\bibitem{butterflies_decline}
M.~S. Warren, D.~Maes, C.~A.~M. van Swaay, P.~Goffart, H.~Van Dyck, N.~A.~D. Bourn, I.~Wynhoff, D.~Hoare, and S.~Ellis.
\newblock The decline of butterflies in europe: Problems, significance, and possible solutions.
\newblock {\em Proceedings of the National Academy of Sciences}, 118(2):e2002551117, 2021.

\bibitem{species_threatened}
{IUCN}.
\newblock The iucn red list of threatened species. version 2024-1, 2024.
\newblock Accessed: 2023-07-14.

\bibitem{bird_and_agri}
S.~J. Ormerod and A.~R. Watkinson.
\newblock Editors' introduction: Birds and agriculture.
\newblock {\em Journal of Applied Ecology}, 37(5):699--705, 2000.

\bibitem{pollinators_decline}
J.~C. Biesmeijer, S.~P.~M. Roberts, M.~Reemer, R.~Ohlemuller, M.~Edwards, T.~Peeters, A.~P. Schaffers, S.~G. Potts, R.~J. M.~C. Kleukers, C.~D. Thomas, et~al.
\newblock Parallel declines in pollinators and insect-pollinated plants in britain and the netherlands.
\newblock {\em Science}, 313(5785):351--354, 2006.

\bibitem{Invasive_species}
R.~P. Keller, J.~Geist, J.~M. Jeschke, and I.~K{\"u}hn.
\newblock Invasive species in europe: ecology, status, and policy.
\newblock {\em Environmental Sciences Europe}, 23:1--17, 2011.

\bibitem{agri_drive_bio_loss}
K.~Henle, D.~Alard, J.~Clitherow, P.~Cobb, L.~Firbank, T.~Kull, D.~McCracken, R.~F.~A. Moritz, J.~Niemel{\"a}, M.~Rebane, et~al.
\newblock Identifying and managing the conflicts between agriculture and biodiversity conservation in europe--a review.
\newblock {\em Agriculture, ecosystems \& environment}, 124(1-2):60--71, 2008.

\bibitem{agri_drive_bio_loss_2}
N.~Dudley and S.~Alexander.
\newblock Agriculture and biodiversity: a review.
\newblock {\em Biodiversity}, 18(2-3):45--49, 2017.

\bibitem{agri_demand_intensification}
T.~K. Rudel, L.~Schneider, M.~Uriarte, B.~L. Turner, R.~DeFries, D.~Lawrence, J.~Geoghegan, S.~Hecht, A.~Ickowitz, E.~F. Lambin, et~al.
\newblock Agricultural intensification and changes in cultivated areas, 1970--2005.
\newblock {\em Proceedings of the national academy of sciences}, 106(49):20675--20680, 2009.

\bibitem{fertilizer_pollution}
S.~Savci.
\newblock An agricultural pollutant: chemical fertilizer.
\newblock {\em International Journal of Environmental Science and Development}, 3(1):73, 2012.

\bibitem{pesticide_biodiversity}
F.~Geiger, J.~Bengtsson, F.~Berendse, W.~W. Weisser, M.~Emmerson, P.~Ceryngier M.~B.~Morales, J.~Liira, T.~Tscharntke, C.~Winqvist, et~al.
\newblock Persistent negative effects of pesticides on biodiversity and biological control potential on european farmland.
\newblock {\em Basic and Applied Ecology}, 11(2):97--105, 2010.

\bibitem{pesticide_biodiversity_2}
A.~Sharma, V.~Kumar, B.~Shahzad, M.~Tanveer, G.P.S. Sidhu, N.~Handa, S.K. Kohli, P.~Yadav, A.S. Bali, R.D. Parihar, and others.
\newblock Worldwide pesticide usage and its impacts on ecosystem.
\newblock {\em SN Applied Sciences}, 1:1--16, 2019.

\bibitem{land_consolidation}
A.~Vitikainen.
\newblock An overview of land consolidation in europe.
\newblock {\em nordic Journal of Surveying and real Estate research}, 1(1), 2004.

\bibitem{land_consolidation_2}
F.~Bartolini and D.~Viaggi.
\newblock The common agricultural policy and the determinants of changes in eu farm size.
\newblock {\em Land use policy}, 31:126--135, 2013.

\bibitem{consolidation_biodiversity}
Y.~Clough, S.~Kirchweger, and J.~Kantelhardt.
\newblock Field sizes and the future of farmland biodiversity in european landscapes.
\newblock {\em Conservation Letters}, 13(6):e12752, 2020.

\bibitem{bio_ecosystem_wellbeing}
R.~Haines-Young, M.~Potschin, et~al.
\newblock The links between biodiversity, ecosystem services and human well-being.
\newblock {\em Ecosystem Ecology: a new synthesis}, 1:110--139, 2010.

\bibitem{bio_ecosystem_wellbeing_2}
P.A. Sandifer, A.E. Sutton-Grier, and B.P. Ward.
\newblock Exploring connections among nature, biodiversity, ecosystem services, and human health and well-being: Opportunities to enhance health and biodiversity conservation.
\newblock {\em Ecosystem services}, 12:1--15, 2015.

\bibitem{conservation_poliocy_EU}
Ian Hodge, Jennifer Hauck, and Aletta Bonn.
\newblock The alignment of agricultural and nature conservation policies in the european union.
\newblock {\em Conservation Biology}, 29(4):996--1005, 2015.

\bibitem{PES_EU}
Boris~T Van~Zanten, Peter~H Verburg, Maria Espinosa, Sergio Gomez-y Paloma, Giuliano Galimberti, Jochen Kantelhardt, Martin Kapfer, Marianne Lefebvre, Rosa Manrique, Annette Piorr, et~al.
\newblock European agricultural landscapes, common agricultural policy and ecosystem services: a review.
\newblock {\em Agronomy for sustainable development}, 34:309--325, 2014.

\bibitem{PES_failure}
David Kleijn, RA~Baquero, Yann Clough, Mario D{\'\i}az, J~De~Esteban, Frederico Fern{\'a}ndez, Doreen Gabriel, Felix Herzog, Andrea Holzschuh, R~J{\"o}hl, et~al.
\newblock Mixed biodiversity benefits of agri-environment schemes in five european countries.
\newblock {\em Ecology letters}, 9(3):243--254, 2006.

\bibitem{multidisciplinary_needed}
N.~Dudley and S.~Alexander.
\newblock Agriculture and biodiversity: a review.
\newblock {\em Biodiversity}, 18(2-3):45--49, 2017.

\bibitem{multidisciplinary_needed_2}
K.~F.~D. Hughey, R.~Cullen, and E.~Moran.
\newblock Integrating economics into priority setting and evaluation in conservation management.
\newblock {\em Conservation Biology}, 17(1):93--103, 2003.

\bibitem{biodiversity_price}
P.~Nijkamp, G.~Vindigni, and P.~A. L.~D. Nunes.
\newblock Economic valuation of biodiversity: A comparative study.
\newblock {\em Ecological economics}, 67(2):217--231, 2008.

\bibitem{biodiversity_price_2}
N.~Hanley and C.~Perrings.
\newblock The economic value of biodiversity.
\newblock {\em Annual Review of Resource Economics}, 11(1):355--375, 2019.

\bibitem{cost-benefit_biodiversity}
G.~Atkinson and S.~Mourato.
\newblock Environmental cost-benefit analysis.
\newblock {\em Annual review of environment and resources}, 33(1):317--344, 2008.

\bibitem{Cost-benefit_critique}
G.~Wegner and U.~Pascual.
\newblock Cost-benefit analysis in the context of ecosystem services for human well-being: A multidisciplinary critique.
\newblock {\em Global environmental change}, 21(2):492--504, 2011.

\bibitem{scarcity_in_eco_price}
M.~A. Drupp, M.~C. H{\"a}nsel, E.~P. Fenichel, M.~Freeman, C.~Gollier, B.~Groom, G.~M. Heal, P.~H. Howard, A.~Millner, F.~C. Moore, et~al.
\newblock Accounting for the increasing benefits from scarce ecosystems.
\newblock {\em Science}, 383(6687):1062--1064, 2024.

\bibitem{Water_scarcity}
F.~Dolan, J.~Lamontagne, R.~Link, M.~Hejazi, P.~Reed, and J.~Edmonds.
\newblock Evaluating the economic impact of water scarcity in a changing world.
\newblock {\em Nature communications}, 12(1):1--10, 2021.

\bibitem{ecosystem_tippingpoing}
V.~Dakos, B.~Matthews, A.P. Hendry, J.~Levine, N.~Loeuille, J.~Norberg, P.~Nosil, M.~Scheffer, and L.~De Meester.
\newblock Ecosystem tipping points in an evolving world.
\newblock {\em Nature ecology \& evolution}, 3(3):355--362, 2019.

\bibitem{bio_price_critique}
E.~G{\'o}mez-Baggethun and M.~Ruiz-P{\'e}rez.
\newblock Economic valuation and the commodification of ecosystem services.
\newblock {\em Progress in physical geography}, 35(5):613--628, 2011.

\bibitem{landsparing_sharing_big_model}
F.~Zabel, R.~Delzeit, J.M. Schneider, R.~Seppelt, W.~Mauser, and T.~V{\'a}clav{\'\i}k.
\newblock Global impacts of future cropland expansion and intensification on agricultural markets and biodiversity.
\newblock {\em Nature communications}, 10(1):2844, 2019.

\bibitem{Mouysset2011}
L.~Mouysset, L.~Doyen, F.~Jiguet, G.~Allaire, and F.~Leger.
\newblock Bio economic modeling for a sustainable management of biodiversity in agricultural lands.
\newblock {\em Ecological Economics}, 70(4):617--626, 2011.

\bibitem{farmers_interaction}
J.~Ahnstr{\"o}m, J.~H{\"o}ckert, H.~Berge{\aa}, C.~Francis, P.~Skelton, and L.~Hallgren.
\newblock Farmers and nature conservation: What is known about attitudes, context factors and actions affecting conservation?
\newblock {\em Renewable agriculture and food systems}, 24(1):38--47, 2009.

\bibitem{farmers_bias}
A.~Thompson, A.~Reimer, and L.~Prokopy.
\newblock Farmers’ views of the environment: the influence of competing attitude frames on landscape conservation efforts.
\newblock {\em Agriculture and human values}, 32:385--399, 2015.

\bibitem{agrilove}
M.~Coronese, M.~Occelli, F.~Lamperti, and A.~Roventini.
\newblock Agrilove: agriculture, land-use and technical change in an evolutionary, agent-based model.
\newblock {\em SSRN Electronic Journal}, 01 2021.

\bibitem{ABM_ecological-economics}
S.~Heckbert, T.~Baynes, and A.~Reeson.
\newblock Agent-based modeling in ecological economics.
\newblock {\em Annals of the new York Academy of Sciences}, 1185(1):39--53, 2010.

\bibitem{farmers_interaction_2}
R.~Greiner.
\newblock Motivations and attitudes influence farmers' willingness to participate in biodiversity conservation contracts.
\newblock {\em Agricultural Systems}, 137:154--165, 2015.

\bibitem{ABM_water_management}
A.~Bourceret, L.~Amblard, and J.D. Mathias.
\newblock Adapting the governance of social--ecological systems to behavioural dynamics: An agent-based model for water quality management using the theory of planned behaviour.
\newblock {\em Ecological Economics}, 194:107338, 2022.

\bibitem{abm_human_nature_interactions}
Li~An.
\newblock Modeling human decisions in coupled human and natural systems: Review of agent-based models.
\newblock {\em Ecological modelling}, 229:25--36, 2012.

\bibitem{sustainable_farming_decision}
Fran{\c{c}}ois~J Dessart, Jes{\'u}s Barreiro-Hurl{\'e}, Ren{\'e} Van~Bavel, et~al.
\newblock Behavioural factors affecting the adoption of sustainable farming practices: a policy-oriented review.
\newblock {\em European Review of Agricultural Economics}, 46(3):417--471, 2019.

\bibitem{abm_land_use}
Robin~B Matthews, Nigel~G Gilbert, Alan Roach, J~Gary Polhill, and Nick~M Gotts.
\newblock Agent-based land-use models: a review of applications.
\newblock {\em Landscape Ecology}, 22:1447--1459, 2007.

\bibitem{agri_ABM_review}
Dimitris Kremmydas, Ioannis~N Athanasiadis, and Stelios Rozakis.
\newblock A review of agent based modeling for agricultural policy evaluation.
\newblock {\em Agricultural systems}, 164:95--106, 2018.

\bibitem{abm_supply_chain}
Dhanan~Sarwo Utomo, Bhakti~Stephan Onggo, and Stephen Eldridge.
\newblock Applications of agent-based modelling and simulation in the agri-food supply chains.
\newblock {\em European Journal of Operational Research}, 269(3):794--805, 2018.

\bibitem{review_ABM_farming_decision}
Robert Huber, Martha Bakker, Alfons Balmann, Thomas Berger, Mike Bithell, Calum Brown, Adrienne Gr{\^e}t-Regamey, Hang Xiong, Quang~Bao Le, Gabriele Mack, et~al.
\newblock Representation of decision-making in european agricultural agent-based models.
\newblock {\em Agricultural systems}, 167:143--160, 2018.

\bibitem{abm_agri_sustainability}
Alba Alonso-Adame, Jef Van~Meensel, Fleur Marchand, Steven Van~Passel, and Siavash Farahbakhsh.
\newblock Sustainability transitions in agri-food systems through the lens of agent-based modeling: a systematic review.
\newblock {\em Sustainability Science}, pages 1--18, 2024.

\bibitem{abm_water_agri_review}
Mohammad~Faiz Alam, Michael McClain, Alok Sikka, and Saket Pande.
\newblock Understanding human--water feedbacks of interventions in agricultural systems with agent based models: a review.
\newblock {\em Environmental Research Letters}, 17(10):103003, 2022.

\bibitem{pampas_bio_ABM}
Diego~O Ferraro, Felipe Ghersa, and Rodrigo Castro.
\newblock Predicting land use and environmental dynamics in argentina's pampas region: An agent-based modeling approach across varied price and climatic scenarios.
\newblock {\em Ecological Modelling}, 498:110881, 2024.

\bibitem{editor_suggestion_2}
J{\"u}rgen Groeneveld, Birgit M{\"u}ller, Carsten~M Buchmann, Gunnar Dressler, Cheng Guo, Niklas Hase, Falk Hoffmann, F~John, Christian Klassert, T~Lauf, et~al.
\newblock Theoretical foundations of human decision-making in agent-based land use models--a review.
\newblock {\em Environmental modelling \& software}, 87:39--48, 2017.

\bibitem{Agripolis}
K.~Happe, K.~Kellermann, and A.~Balmann.
\newblock Agent-based analysis of agricultural policies: an illustration of the agricultural policy simulator agripolis, its adaptation and behavior.
\newblock {\em Ecology and society}, 11(1), 2006.

\bibitem{abm_heuristic_weed}
Robert Huber, Hang Xiong, Kevin Keller, and Robert Finger.
\newblock Bridging behavioural factors and standard bio-economic modelling in an agent-based modelling framework.
\newblock {\em Journal of Agricultural Economics}, 73(1):35--63, 2022.

\bibitem{ABM_criticism}
G.~Fagiolo, A.~Moneta, and P.~Windrum.
\newblock A critical guide to empirical validation of agent-based models in economics: Methodologies, procedures, and open problems.
\newblock {\em Computational Economics}, 30:195--226, 2007.

\bibitem{JP_economic_complexity_manifesto}
Jean-Philippe Bouchaud.
\newblock Navigating through economic complexity: Phase diagrams \& parameter sloppiness, 2024.

\bibitem{mark0}
Stanislao Gualdi, Marco Tarzia, Francesco Zamponi, and Jean-Philippe Bouchaud.
\newblock Tipping points in macroeconomic agent-based models.
\newblock {\em Journal of Economic Dynamics and Control}, 50:29--61, 2015.

\bibitem{paz_habitat_fragmentation}
Diego Bengochea~Paz, Kirsten Henderson, and Michel Loreau.
\newblock Habitat percolation transition undermines sustainability in social-ecological agricultural systems.
\newblock {\em Ecology Letters}, 25(1):163--176, 2022.

\bibitem{ODD_grim_2020}
Volker Grimm, Steven~F Railsback, Christian~E Vincenot, Uta Berger, Cara Gallagher, Donald~L DeAngelis, Bruce Edmonds, Jiaqi Ge, Jarl Giske, Juergen Groeneveld, et~al.
\newblock The odd protocol for describing agent-based and other simulation models: A second update to improve clarity, replication, and structural realism.
\newblock {\em Journal of Artificial Societies and Social Simulation}, 23(2), 2020.

\bibitem{ODD}
Birgit M{\"u}ller, Friedrich Bohn, Gunnar Dre{\ss}ler, J{\"u}rgen Groeneveld, Christian Klassert, Romina Martin, Maja Schl{\"u}ter, Jule Schulze, Hanna Weise, and Nina Schwarz.
\newblock Describing human decisions in agent-based models--odd+ d, an extension of the odd protocol.
\newblock {\em Environmental Modelling \& Software}, 48:37--48, 2013.

\bibitem{share_rent_UAA}
P.~Ciaian, D.~Kancs, J.~Swinnen, K.~Van Herck, and L.~Vranken.
\newblock {\em Key Issues and Developments in Farmland Rental Markets in EU Member States and Candidate Countries}.
\newblock Centre for European Policy Studies (CEPS), Brussels (Belgium), 2012.

\bibitem{bird_as_good_index}
European~Committee of~the Regions, Climate~Change Commission for~the Environment, Energy, M.~Gancheva, S.~O’Brien, C.~Moreno, and A.~Valentino.
\newblock {\em Towards an 8th Environment Action Programme – Local and regional dimension}.
\newblock European Committee of the Regions, 2018.

\bibitem{Loreau_agri_ecosystem}
D.~Montoya, B.~Haegeman, s.~Gaba, C.~de~Mazancourt, V.~Bretagnolle, and M.~Loreau.
\newblock Trade-offs in the provisioning and stability of ecosystem services in agroecosystems.
\newblock {\em Ecological Applications}, 29(2):e01853, 2019.

\bibitem{Loreau_agri_ecosystem_3}
D.~Montoya, B.~Haegeman, s.~Gaba, C.~de~Mazancourt, V.~Bretagnolle, and M.~Loreau.
\newblock Habitat fragmentation and food security in crop pollination systems.
\newblock {\em Journal of Ecology}, 109(8):2991--3006, 2021.

\bibitem{bird_and_pest_4}
Christopher~J Whelan, Daniel~G Wenny, and Robert~J Marquis.
\newblock Ecosystem services provided by birds.
\newblock {\em Annals of the New York academy of sciences}, 1134(1):25--60, 2008.

\bibitem{bird_and_pest}
P.~Díaz-Siefer, N.~Olmos-Moya, F.~Fontúrbel, B.~Lavandero, R.~Pozo, and J.~Celis-Diez.
\newblock Bird-mediated effects of pest control services on crop productivity: a global synthesis.
\newblock {\em Journal of Pest Science}, 95, 03 2022.

\bibitem{bird_and_pest_2}
Lushka Labuschagne, Lourens~H Swanepoel, Peter~J Taylor, Steven~R Belmain, and Mark Keith.
\newblock Are avian predators effective biological control agents for rodent pest management in agricultural systems?
\newblock {\em Biological Control}, 101:94--102, 2016.

\bibitem{bird_and_pest_3}
Lian~Pin Koh.
\newblock Birds defend oil palms from herbivorous insects.
\newblock {\em Ecological applications}, 18(4):821--825, 2008.

\bibitem{FADN}
{European Commission, Directorate-General for Agriculture and Rural Development}.
\newblock Farm accountancy data network public database, 2019.
\newblock Accessed: 2024-07-10.

\bibitem{Mitcherliche_prod_func}
Charles~B Moss.
\newblock {\em Production Economics}.
\newblock World Scientific, 2022.

\bibitem{future_food_demand}
A.~Tibi, A.~Forslund, P.~Debaeke, B.~Schmitt, H.~Guyomard, E.~Marajo-Petitzon, T.~Ben-Ari, A.~Bérard, A.~Bispo, J.L. Durand, P.~Faverdin, J.~Le Gouis, D.~Makowski, and S.~Planton.
\newblock Place des agricultures européennes dans le monde à l’horizon 2050 : entre enjeux climatiques et défis de la sécurité alimentaire. rapport de synthèse de l'étude.
\newblock Rapport de synthèse, INRAE, France, 2020.
\newblock + Annexes.

\bibitem{economy_of_scale}
M.~Duffy.
\newblock Economies of size in production agriculture.
\newblock {\em Journal of hunger \& environmental nutrition}, 4:375--392, 07 2009.

\bibitem{economy_of_scale_2}
F.~Kuhlmann and E.~Berg.
\newblock The farm as an enterprise--the european perspective.
\newblock 2002.

\bibitem{CAP_no_price_support}
OECD.
\newblock {\em Evaluation of Agricultural Policy Reforms in the European Union}.
\newblock 2011.

\bibitem{pesticide_efficiency}
J.~Xiao, L.~Chen, F.~Pan, Y.~Deng, C.~Ding, M.~Liao, X.~Su, and H.~Cao.
\newblock Application method affects pesticide efficiency and effectiveness in wheat fields.
\newblock {\em Pest Management Science}, 76(4):1256--1264, 2020.

\bibitem{tech_ABM}
G.~Dosi.
\newblock Technological paradigms and technological trajectories: A suggested interpretation of the determinants and directions of technical change.
\newblock {\em Research Policy}, 11(3):147--162, 1982.

\bibitem{land_market}
K.~Kellermann, C.~Sahrbacher, and A.~Balmann.
\newblock Land markets in agent based models of structural change.
\newblock 02 2008.

\bibitem{pesticide_app_behaviour}
E.~Meunier, P.~Smith, T.~Griessinger, and C.~Robert.
\newblock Understanding changes in reducing pesticide use by farmers: Contribution of the behavioural sciences.
\newblock {\em Agricultural Systems}, 214:103818, 2024.

\bibitem{adaptation_farming_decision}
Marion Robert, Alban Thomas, and Jacques-Eric Bergez.
\newblock Processes of adaptation in farm decision-making models. a review.
\newblock {\em Agronomy for sustainable development}, 36:1--15, 2016.

\bibitem{DSK_macro_ABM}
Francesco Lamperti, Giovanni Dosi, Mauro Napoletano, Andrea Roventini, and Alessandro Sapio.
\newblock Faraway, so close: Coupled climate and economic dynamics in an agent-based integrated assessment model.
\newblock {\em Ecological Economics}, 150:315--339, 2018.

\bibitem{bird_logistic}
L.~Mouysset, M.~Miglianico, D.~Makowski, F.~Jiguet, and D.~Luc.
\newblock Selection of dynamic models for bird populations in farmlands.
\newblock {\em Environmental Modeling \& Assessment}, 21, 06 2016.

\bibitem{dhondt1988carrying}
A.~Dhondt.
\newblock Carrying capacity: a confusing concept.
\newblock {\em ACTA OECOL.(OECOL. GEN.).}, 9(4):337--346, 1988.

\bibitem{yield_biodiversity}
B.~J. Cardinale, C.~T. Harvey, K.~Gross, and A.~R. Ives.
\newblock Biodiversity and biocontrol: emergent impacts of a multi-enemy assemblage on pest suppression and crop yield in an agroecosystem.
\newblock {\em Ecology letters}, 6(9):857--865, 2003.

\bibitem{Bird_index_data}
European Bird~Census Council, Royal~Society for the Protection~of Birds, BirdLife International, and Czech~Society for Ornithology.
\newblock Geographical coverage: Eu=eu-27 (except malta), 2008.
\newblock Source provided by European Bird Census Council, Royal Society for the Protection of Birds, BirdLife International, and Czech Society for Ornithology.

\bibitem{faostructuraldata}
{Food and Agriculture Organization of the United Nations (FAO)}.
\newblock Structural data from agricultural censuses.
\newblock FAOSTAT database, Accessed 2024.
\newblock Accessed on July 10, 2024.

\bibitem{FAO_price}
{Food and Agriculture Organization of the United Nations (FAO)}.
\newblock Agriculture producer prices and producer price index.
\newblock FAOSTAT database, Accessed 2024.
\newblock Accessed on July 10, 2024.

\bibitem{potential_pest_damage_report_EU}
European Parliament, Directorate-General for Parliamentary Research~Services, D~Bylemans, B~De~Coninck, and W~Keulemans.
\newblock {\em Farming without plant protection products – Can we grow without using herbicides, fungicides and insecticides?}
\newblock Publications Office, 2019.

\bibitem{maximum_yield}
R.~Schils, J.E. Olesen, K.-C. Kersebaum, B.~Rijk, M.~Oberforster, V.~Kalyada, M.~Khitrykau, A.~Gobin, H.~Kirchev, V.~Manolova, et~al.
\newblock Cereal yield gaps across europe.
\newblock {\em European Journal of Agronomy}, 101:109--120, 2018.

\bibitem{yield_noise}
B.~Joernsgaard and S.~Halmoe.
\newblock Intra-field yield variation over crops and years.
\newblock {\em European Journal of Agronomy}, 19(1):23--33, 2003.

\bibitem{pesticide_price}
European Parliament, Directorate-General for Parliamentary Research~Services, J~Bremmer, M~Manshanden, A~Smit, and J~Jager.
\newblock {\em Cost of crop protection measures – A follow-up to the study 'The future of crop protection in Europe' (2021)}.
\newblock European Parliament, 2021.

\bibitem{EU_Cereal_costs}
European Commission.
\newblock Eu cereal farms report: Based on 2017 fadn data, 2017.
\newblock Accessed: 2024-07-13.

\bibitem{CAP_history}
A.~Giuliani and H.~Baron.
\newblock The cap (common agricultural policy): A short history of crises and major transformations of european agriculture.
\newblock In {\em Forum for Social Economics}, pages 1--27. Taylor \& Francis, 2023.

\bibitem{Farm_to_Fork}
J.~Wesseler.
\newblock The eu's farm-to-fork strategy: An assessment from the perspective of agricultural economics.
\newblock {\em Applied Economic Perspectives and Policy}, 44(4):1826--1843, 2022.

\bibitem{pesticide_tax}
T.~B{\"o}cker and R.~Finger.
\newblock European pesticide tax schemes in comparison: an analysis of experiences and developments.
\newblock {\em Sustainability}, 8(4):378, 2016.

\bibitem{organic_EU}
M.~Stolze, A.~Piorr, A.M. H{\"a}ring, and S.~Dabbert.
\newblock {\em Environmental impacts of organic farming in Europe}.
\newblock Universit{\"a}t Hohenheim, Stuttgart-Hohenheim, 2000.

\bibitem{yield_reduction_by_pesticide_reduction_EU_report}
European Commission.
\newblock Commission response to council decision (eu) 2022/2572 of 19 december 2022, 2023.
\newblock Accessed: 2023-07-14.

\bibitem{pesticide_reduction_policy}
Niklas M{\"o}hring, Karin Ingold, Per Kudsk, Fabrice Martin-Laurent, Urs Niggli, Michael Siegrist, Bruno Studer, Achim Walter, and Robert Finger.
\newblock Pathways for advancing pesticide policies.
\newblock {\em Nature food}, 1(9):535--540, 2020.

\bibitem{EU_biodiversity_plan}
European Commission and Directorate-General for Environment.
\newblock {\em EU biodiversity strategy for 2030 – Bringing nature back into our lives}.
\newblock Publications Office of the European Union, 2021.

\bibitem{pesticide_human_health}
A.F. Hern{\'a}ndez, T.~Parr{\'o}n, A.M. Tsatsakis, M.~Requena, R.~Alarc{\'o}n, and O.~L{\'o}pez-Guarnido.
\newblock Toxic effects of pesticide mixtures at a molecular level: their relevance to human health.
\newblock {\em Toxicology}, 307:136--145, 2013.

\bibitem{bio_policy_beyond_economic}
Iago Otero, Katharine~N Farrell, Salvador Pueyo, Giorgos Kallis, Laura Kehoe, Helmut Haberl, Christoph Plutzar, Peter Hobson, Jaime Garc{\'\i}a-M{\'a}rquez, Beatriz Rodr{\'\i}guez-Labajos, et~al.
\newblock Biodiversity policy beyond economic growth.
\newblock {\em Conservation letters}, 13(4):e12713, 2020.

\bibitem{stechemesser2024climate}
A.~Stechemesser, N.~Koch, E.~Mark, E.~Dilger, P.~Kl{\"o}sel, L.~Menicacci, Nachtigall, et~al.
\newblock Climate policies that achieved major emission reductions: Global evidence from two decades.
\newblock {\em Science}, 385(6711):884--892, 2024.

\bibitem{different_farm_systems}
Christian Bockstaller, Laurence Guichard, David Makowski, Anne Aveline, Philippe Girardin, and Sylvain Plantureux.
\newblock Agri-environmental indicators to assess cropping and farming systems. a review.
\newblock {\em Agronomy for sustainable development}, 28:139--149, 2008.

\bibitem{different_farm_systems_2}
Tiziano Gomiero, David Pimentel, and Maurizio~G Paoletti.
\newblock Environmental impact of different agricultural management practices: conventional vs. organic agriculture.
\newblock {\em Critical reviews in plant sciences}, 30(1-2):95--124, 2011.

\bibitem{risk_farming_decision}
Caroline Roussy, Aude Ridier, and Karim Chaib.
\newblock Farmers' innovation adoption behaviour: role of perceptions and preferences.
\newblock {\em International Journal of Agricultural Resources, Governance and Ecology}, 13(2):138--161, 2017.

\bibitem{semi_natural_biodiversity}
John~M Holland, Jacob~C Douma, Liam Crowley, Laura James, Laura Kor, David~RW Stevenson, and Barbara~M Smith.
\newblock Semi-natural habitats support biological control, pollination and soil conservation in europe. a review.
\newblock {\em Agronomy for Sustainable Development}, 37:1--23, 2017.

\bibitem{Sobol}
I.M. Sobol.
\newblock On the distribution of points in a cube and the approximate evaluation of integrals.
\newblock {\em Ussr Computational Mathematics and Mathematical Physics}, 7:86--112, 1967.

\bibitem{R2adjusted}
H.C. Carver.
\newblock Reviewed work: Methods of correlation analysis by mordecai ezekiel.
\newblock {\em Journal of the American Statistical Association}, 26(175):350--353, 1931.

\end{thebibliography}

\newpage

\appendix

\section{Calibration procedure and sensitivity analysis} 
\label{app:calibration_sensitivity}

In order to fine-tune the model with historical data, we employed a simple sampling-based approach. The hypothesized ranges for each calibration parameter are detailed in Table~\ref{tab:calibration}. To comprehensively explore this parameter space, we generated 4096 sample points using a Sobol sequence~\cite{Sobol}. For each parameter combination, we ran 10 Monte Carlo simulations, and computed  the sum of squared differences between simulated and observed data. To ensure an equal number of data points in the different observational time series, we interpolated linearly the structural data (farmer count, land utilization, farm size), consistent with the observed linear trends. Table~\ref{tab:calibration} displays the results. The corresponding R-squared $R^2$ and adjusted R-squared $\bar R^2$~\cite{R2adjusted} values demonstrate that the model effectively captures the observed dynamics in real data.

\begin{table}[ht]
    \centering
    \begin{tabular}{lccc}
        \toprule
         & Description & Sampling range & Optimal \\
        \midrule
        $\alpha$ & Price frictions & [0.01, 0.1] & 0.08 \\
        $\lambda$ & Yield target adjustment speed & [0.1, 0.5] & 0.2 \\
        $\gamma$ & Pesticide adjustment speed  & [1.0, 3.5] & 3.0 \\
        $\beta$ & Land adjustment speed & [0.4, 0.5] & 0.45 \\
        $\eta$ & Profit share for technology & [0.05, 0.5] & 0.15 \\
        $\mathcal{P}_{\mathrm{ref}}$ & Reference profit for technology & [150, 1500] & 1000.0 \\
        $\nu_{\max}$ & Maximum efficiency gain & [0.05, 0.5] & 0.10 \\
        \hdashline[.4pt/1pt]
        $\overline{\Delta e}$ & Average efficiency gain  &  & 0.025 \\
        $R^2$ & R-squared &  & 0.719 \\
        $\bar{R}^2$ & Adjusted R-squared &  & 0.716 \\
        \bottomrule
    \end{tabular}
    \caption{Results of model calibration. The resulting average efficiency gain per year across all farmers $\overline{\Delta e}$ is not a parameter; it is computed from the resulting time series of each simulation.}
    \label{tab:calibration}
\end{table}

\begin{figure}[htbp] 
\centering
\includegraphics[width=\textwidth]{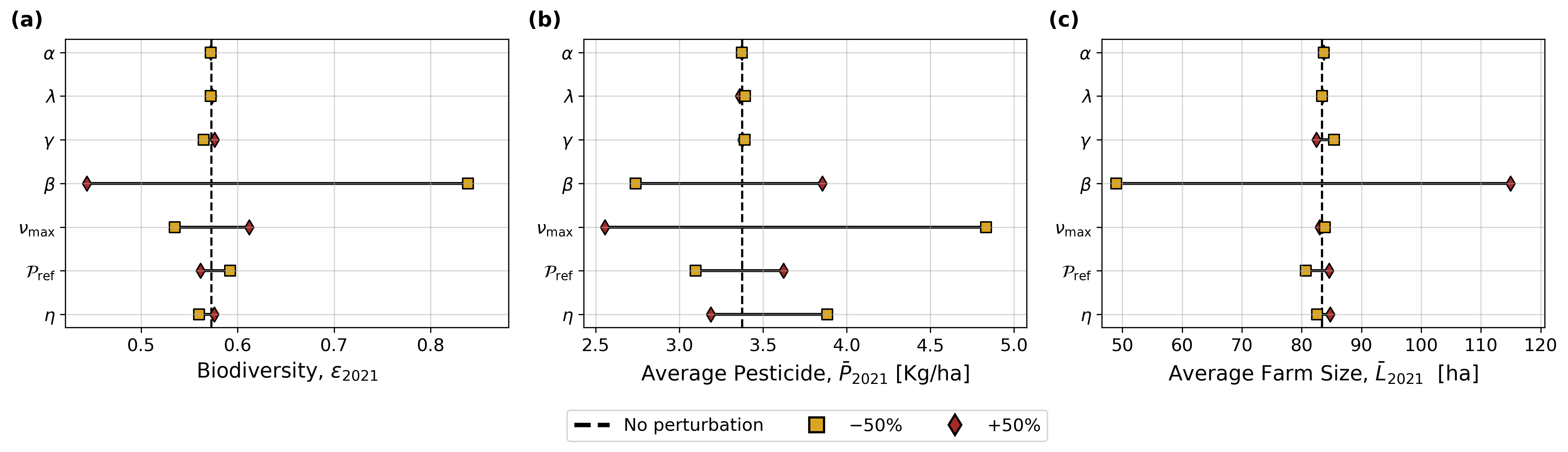}
\caption{Sensitivity analysis for the calibrated parameters (see Table~\ref{tab:params}). This figure shows the variation in three main variables $\varepsilon_t, \, \bar P_t$ and $\bar L_t$ in the year 2020 when the values in Table~\ref{tab:params} are independently perturbed by ±50\%. As expected, the parameter $\beta$, which determines the land speed update, exerts the most significant influence on the dynamics across all variables. 
}
\label{fig:sensitivity}
\end{figure}

{To assess the model's sensitivity to parameter variations, each calibrated parameter was independently perturbed by ±50\%. The resulting impacts on the three key variables, $\varepsilon_t$, $\bar{P}_t$, and $\bar{L}_t$, are illustrated in Fig.~\ref{fig:sensitivity}. Our analysis reveals that the land adjustment speed parameter, $\beta$, exerts the most significant influence on the model's dynamics, as evidenced by the wide range of variation observed. This substantial influence underscores the critical role of land consolidation phenomena, which are heavily dependent on the value of $\beta$. In contrast, the parameters related to technology ($\nu_{\max}$, $\mathcal{P}_{\mathrm{ref}}$, and $\eta$) are particularly relevant in determining pesticide adoption rates, while exhibiting a limited impact on biodiversity. This limited impact can be attributed to the high value of $\mu$, which effectively limits the impact of pesticide use on biodiversity evolution. Finally, the adjustment speeds  ($\alpha$, $\lambda$, and $\gamma$) exhibit less variability, suggesting that they have a relatively minor impact on the system's dynamics. This observed robustness in the model's results, highlighting the minimal impact of specific behavioral details, addresses one of the primary criticisms often leveled against agent-based models~\cite{ABM_criticism}. }

\section{{Biodiversity influence on yield}}
\label{app:bio_yield}

\begin{figure}[h!]
\centering
\includegraphics[width=\textwidth]{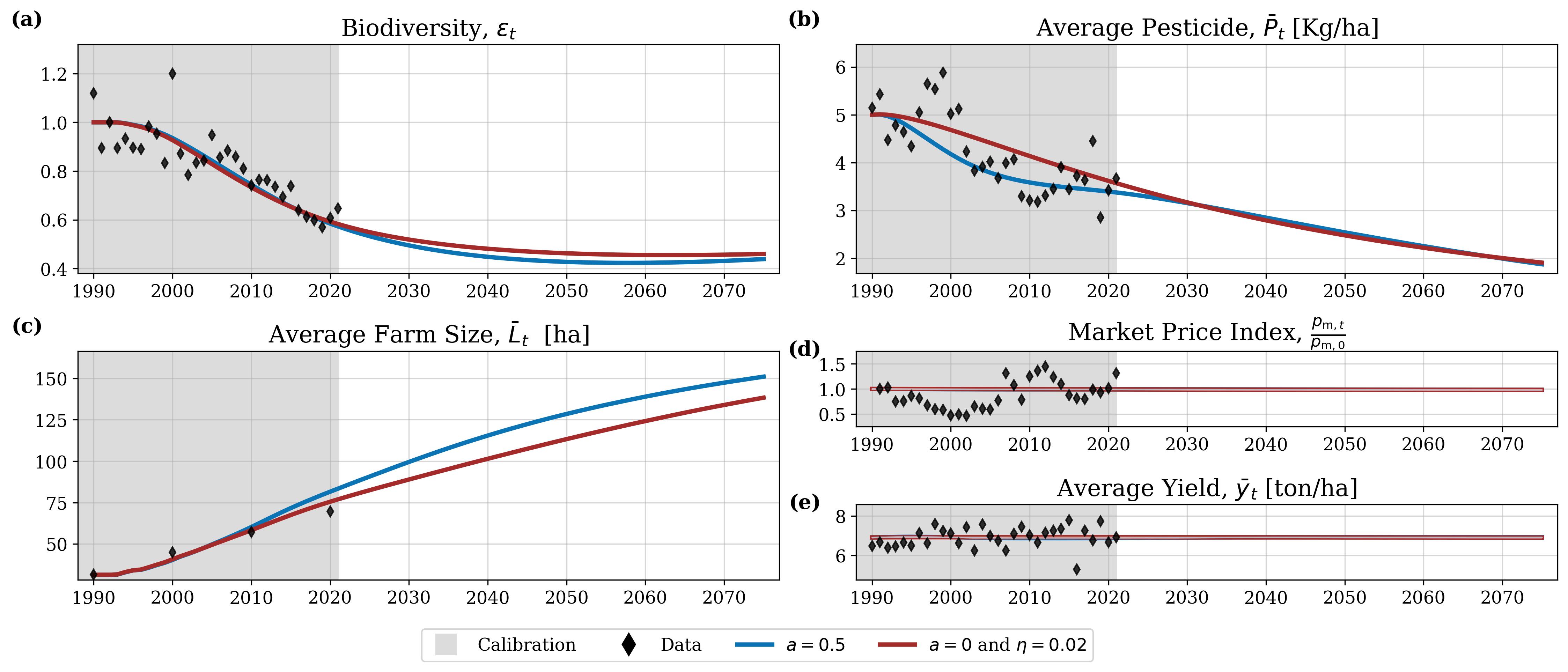}
\caption{Effect of biodiversity on pest exposure and yield. This figure compares model outcomes for two scenarios: (1) the baseline scenario where pest exposure is influenced by biodiversity ($a=0.5$), and (2) a scenario where pest exposure has no dependency on biodiversity ($a=0$). The parameter $\eta$, controlling the fraction of profit devoted to technology adoption, has been recalibrated to fit historical trends. Results show that reducing the influence of biodiversity on pest control (scenario $a=0$) leads to reduced land consolidation patterns and thus slows the rate of biodiversity decline. While these results temper the model's conclusions, the overall message remains valid: a policy combining both subsidies and pesticide reduction is necessary to halt biodiversity decline.}
\label{fig:biodiversity_pest_impact}
\end{figure}

Although it is widely recognized that higher levels of biodiversity contribute to natural pest control and thereby enhance crop productivity, the extent of this effect remains a topic of debate~\cite{bird_and_pest_4, bird_and_pest, bird_and_pest_2, bird_and_pest_3}. In this section, we examine the consequences of diminishing the impact of biodiversity on crop yield by altering the parameter $a$ in Eq.~\ref{eq:pest}. Specifically, by decreasing $a$, we decrease the dependency of yield on biodiversity, making pest exposure less reliant on it. To evaluate the impact of varying $a$ and compare it to the baseline results, we adjusted the parameter $\eta$, which represents the profit share devoted to technological improvements. This adjustment reflects the reduced need for pest management investment when pest exposure is lessened. The results  are presented in Fig~\ref{fig:biodiversity_pest_impact}, in particular for with $a=0$.

The overall decline in biodiversity remains consistent with the scenario where $a=0.5$, but the rate of decline is moderated in the absence of the feedback loop. With decreased pest exposure, there is less focus on research and development efforts and a reduced allocation of funds (where $\eta = 0.02$). Consequently, farmers experience increased income, which helps mitigating the land consolidation trend.

\section{{Technological advances and sustainable farming}}\label{app:tech}

To incorporate the impact of technological progress, particularly advancements that may lead to less sustainable farming practices such as the use of more potent pesticides, we modify Eq.~\ref{eq:car_cap} by including the technological contribution within the weighted average of pesticide effects:

\begin{align}
K_{t+1} = \mu\frac{\bar{L}_0}{\bar{L}_{t+1}} + (1-\mu) \frac{\sum_{i=1}^{N_0} L_{i,0} P_{i,0} (e_{i,0})^k}{\sum_{i=1}^{N_0} L_{i,0}} \frac{\sum_{i=1}^{N_{t+1}} L_{i,t}}{\sum_{i=1}^{N_{t+1}} L_{i,{t+1}}  P_{i,{t+1}} (e_{i,{t+1}})^k }.
\end{align}

\begin{figure}[t!]
\centering
\includegraphics[width=\textwidth]{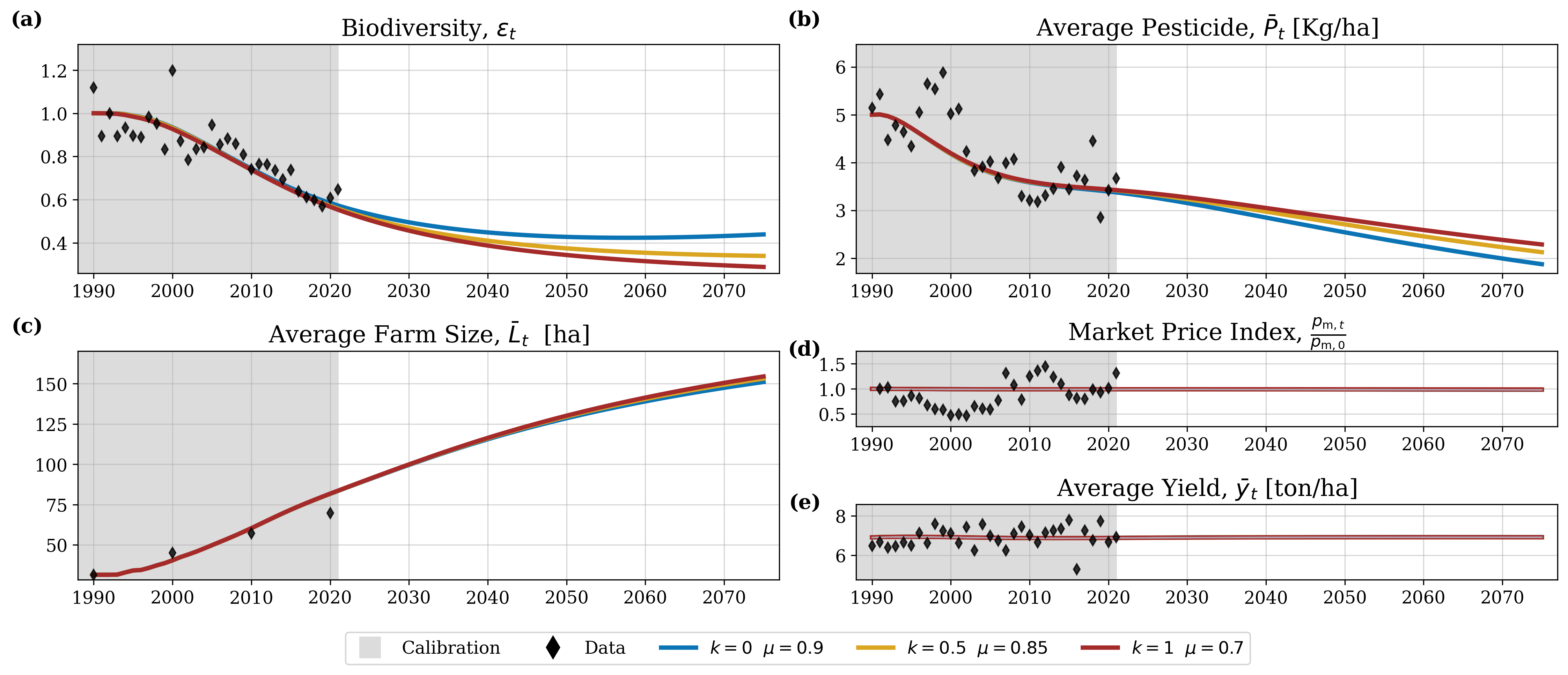}
\caption{Impact of technological advancement on biodiversity decline. This figure illustrates the effects of incorporating technologies that do not contribute to reducing the impact of pesticide use on biodiversity by varying the parameter $k$. In particular, increasing the parameter $k$ corresponds to neglecting the positive effects of pesticide reduction. The extreme case of $k=1$ represents a scenario where technological progress offers no biodiversity benefits. For each value of $k$, we re-estimated the value of $\mu$ to fit historical trends. Results show that as $k$ increases, there is a notable acceleration in biodiversity decline. This deterioration is not solely due to increased pesticide use; it is also exacerbated by intensified land consolidation. Indeed, to maintain the same yield, there is an increased reliance on technology, which small farmers often struggle to compete with. This highlights the importance of supporting small farmers while reducing pesticide usage, thereby strengthening the model results.}
\label{fig:different_k}
\end{figure}

In this modified framework, when $k=0$, the previous results are retained. However, for $k>0$, technological advancements contribute to biodiversity decline. The extreme case of $k=1$ represents a scenario where technological progress offers no biodiversity benefits. To evaluate the impact of varying $k$ and compare it to the baseline results, we re-estimated the parameter $\mu$ to fit historical trends. As shown in Fig~\ref{fig:different_k}, a higher value of $k$ necessitates a lower $\mu$ value, indicating a greater influence of pesticide use on biodiversity decline.

Upon examining forecasted biodiversity, it is evident that increasing $k$ correlates with a general decline in biodiversity, consistent with expectations. Notably, this deterioration is not solely attributed to increased pesticide use but is also exacerbated by heightened land consolidation. As detailed in the main text, the loss of biodiversity disproportionately affects small farmers, who find it challenging to compete with the required technology. These findings suggest that policies supporting small farmers while reducing pesticide use are becoming increasingly critical.

These insights highlight the necessity for a nuanced analysis of efficiency in agricultural systems and the role of green technology adoption by farmers in shaping biodiversity outcomes. However, the interplay of technological adoption, land consolidation, and behavioral dynamics requires a sophisticated approach that considers these multiple factors, as elaborated in \cite{agrilove}.

\end{document}